\documentclass[aps,prc,twocolumn,superscriptaddress,showpacs,preprintnumbers,nofootinbib]{revtex4-1}
\usepackage{epsfig,bm,epsf,float,footnote}
\usepackage{amsmath}
\usepackage{amsthm}
\usepackage{mathrsfs}
\usepackage{soul}
\usepackage[normalem]{ulem}
\usepackage[dvipdfmx]{color}
\usepackage[breaklinks=true]{hyperref}

\begin{document}

\preprint{YITP-17-06}

\title{Few-body approach to structure of $\bar{K}$-nuclear quasi-bound states}


\author{Shota Ohnishi}
\affiliation{Department of Physics, Hokkaido University, Sapporo 060-0810, Japan}

\author{Wataru Horiuchi}
\affiliation{Department of Physics, Hokkaido University, Sapporo 060-0810, Japan}

\author{Tsubasa Hoshino}
\affiliation{Department of Physics, Hokkaido University, Sapporo 060-0810, Japan}

\author{Kenta Miyahara}
\affiliation{Department of Physics, Kyoto University,
Kyoto 606-8502, Japan}

\author{Tetsuo Hyodo}
\affiliation{Yukawa Institute for Theoretical Physics, Kyoto University, Kyoto 606-8502, Japan}


\date{\today}

\begin{abstract}
  Structure of
  light antikaon-nuclear quasi-bound states, which consist
    of an antikaon $(\bar{K}=K^-,~\bar{K}^0)$
and a few nucleons $(N=p,~n)$ such as
$\bar{K}NN$, $\bar{K}NNN$,
$\bar{K}NNNN$ and $\bar{K}NNNNNN$ systems, is studied
with full three- to seven-body calculations.
Employing a realistic $\bar{K}N$ potential based on the chiral SU(3)
effective field theory with the SIDDHARTA constraint, we show that the central nucleon densities of these systems increases when the antikaon is injected, by about factor of two at maximum.
The $\bar{K}NNNN$ system shows the largest central density, about 0.74 fm$^{-3}$ even with the phenomenological $\bar{K}N$ potential, which are not as high as those suggested in previous studies with approximate treatments of the few-body systems.
  We find the spin of the ground state of the $\bar{K}NNNNNN$ system
depends on the strength of the $\bar{K}N$ attraction. Thus, the quantum number of the ground state can be another constraint on the $\bar{K}N$ interaction.
\end{abstract}


\maketitle

\section{Introduction}

In recent years, properties of the antikaon($\bar{K}$)-nuclear quasi-bound
states, so-called kaonic nuclei, have been studied actively.
Since the nominal location of the $\Lambda(1405)$ mass is slightly below
the $K^-p$ threshold~\cite{PDG2016},
the $\Lambda(1405)$ is considered as a $\bar{K}N$ quasi-bound state embedded in
the $\pi\Sigma$ continuum~\cite{Dalitz:1959dn,Dalitz:1960du}.
Motivated by such a picture, phenomenological $\bar{K}N$ interaction
models were constructed 
so that they
reproduce the $\Lambda(1405)$ nominal mass together
with two-body scattering data~\cite{Akaishi:2002bg,Shevchenko:2011ce}.
The strong attraction of the phenomenological potential models
  predicts deeply-bound $\bar{K}$ states in light nuclei with
  binding energy larger than $100$ MeV, and extremely
  dense systems
  about ten times higher than the ordinary nuclear density~\cite{Akaishi:2002bg,Yamazaki:2002uh,Dote:2002db,Dote:2003ac}.
  It should, however, be noted that the few-body problem was not accurately solved to predict such high-density systems, but the optical potential model
  or the $g$-matrix approach were adopted. The validity of those approaches should be examined with care, at least in the few-body systems.

The $\bar{K}N$ interactions are essential for 
determining the structure of the kaonic nuclei.
The $\bar{K}$ 
belongs to a part of the pseudscalar octet
of Nambu-Goldstone bosons associated
with the spontaneous symmetry breaking of chiral
SU(3)${}_L\times$SU(3)${}_R$ in low-energy QCD.
Thus, the chiral SU(3) effective field theory based on the symmetry breaking mechanism is a more systematic framework to obtain the $\bar{K}N$ interaction, and has
 succeeded in dealing with the
$\bar{K}N$ interaction with $\bar{K}N$-$\pi\Sigma$ couplings~\cite{Kaiser:1995eg,Oset:1997it,Oller:2000fj,Hyodo:2011ur}.
In fact, including the next-to-leading order (NLO) contributions, the chiral SU(3) approach reproduces all existing experimental data at the level of $\chi^{2}/$d.o.f $\sim 1$~\cite{Ikeda:2011pi,Ikeda:2012au}. Among others, the precise measurement of the kaonic hydrogen by the SIDDHARTA collaboration~\cite{Bazzi:2011zj,Bazzi:2012eq} gives strong constraint at the $\bar{K}N$ threshold, with which the uncertainty in the subthreshold extrapolation of the $\bar{K}N$ amplitude is significantly reduced. 
The equivalent single-channel $\bar{K}N$ potential to the NLO chiral dynamics including the SIDDHARTA constraint is constructed in Ref.~\cite{Miyahara:2015bya} based on the framework presented in Ref.~\cite{Hyodo:2007jq}. Thus, the realistic $\bar{K}N$ potential is now available.

The $\bar{K}N$-$\pi\Sigma$ scattering amplitude from the chiral SU(3)
dynamics has two poles in the $\Lambda(1405)$ energy region~\cite{Oller:2000fj,Jido:2003cb,Minireview}: one is located around $1420$
MeV, while the other exhibits a
broad resonant structure above the $\pi\Sigma$ threshold.
The pole located around $1420$ MeV corresponds
to the $\bar{K}N$ quasi-bound state
with the binding energy of $15$ MeV, about a half of the
binding energy
assumed in the phenomenological $\bar{K}N$ interactions.
This different pole structure comes from different off-shell properties
of the $\bar{K}N$ interactions.
The $\bar{K}N$ interaction based on the chiral SU(3) dynamics is
energy-dependent, and that in the subthreshold becomes
less attractive than the one proposed by the
energy-independent phenomenological potential~\cite{Hyodo:2007jq}.
These different off-shell properties also appear in how the $\Lambda(1405)$
resonance shows up in the differential cross
section of the $K^-d\rightarrow \pi\Sigma n$ reaction~\cite{Ohnishi:2015iaq}.
These differences are further enhanced in the light kaonic nuclei.
For the lightest kaonic nuclei so-called strange dibaryons in the $\bar{K}NN$-$\pi YN$ ($Y=\Sigma$,~$\Lambda$) coupled system,
the energy-dependent potential
models~\cite{Dote:2008in,Dote:2008hw,Ikeda:2010tk,Barnea:2012qa,Dote:2014via}
give resonance energies higher
than the energy-independent
ones~\cite{Yamazaki:2002uh,Yamazaki:2007cs,Wycech:2008wf,Shevchenko:2006xy,Shevchenko:2007ke,Ikeda:2007nz,Ikeda:2008ub}.
How a possible signature of this strange dibaryon resonance shows up in
the resonance production reaction is also of interest as it reflects the
two-body dynamics of the $\bar{K}N$ system~\cite{Ohnishi:2013rix}.

Given the background described above, we raise three questions to be discussed in this paper; 1) What are the structure of light kaonic nuclei when the reliable $NN$ and $\bar{K}N$ interactions are used? 2) Can the high-density $\bar{K}$ nuclei be realized within the accurate few-body treatment? 3) How the off-shell dependence of the $\bar{K}N$ interaction affect the few-body systems? To answer these questions, we perform
fully microscopic few-body calculations
for three- to seven-body systems including an antikaon.
Here, the systems with a $\bar{K}$ and $(\mathscr{N}-1)$ nucleons
are accurately described by employing the
stochastic variational method (SVM) with correlated Gaussian (CG)
basis~\cite{Varga:1995dm,Varga:1997xga,suzuki1998stochastic}.
We employ the $\bar{K}N$ interaction based on the chiral SU(3) dynamics with the SIDDHARTA constraint~\cite{Miyahara:2015bya} as a realistic $\bar{K}N$ force. Combining them with the reliable nuclear forces, we  
present quantitative predictions of the structure of the light kaonic nuclei.
Next, we perform the same few-body calculations with the phenomenological $\bar{K}N$ interaction, so-called Akaishi-Yamazaki (AY) potential~\cite{Akaishi:2002bg,Yamazaki:2007cs}, in order to examine the validity of the many-body approximations used in the prediction of the high-density states.
Furthermore, the comparison of the results with two $\bar{K}N$ potentials 
serves as a study of the off-shell dependence of the interactions.
In this way, we 
systematically study the structure of kaonic nuclei
and 
discuss how the nuclear structure is changed by $\bar{K}$.

In Sec.~\ref{sec:two-body},
we briefly review
the two-body interactions used in this work.
The SVM with the CG
for the $\mathscr{N}$-body systems is 
explained in Sec.~\ref{sec:SVM}.
We summarize the quantities to analyze the structures of the few-body systems in Sec.~\ref{sec:structure}.
Numerical results of the properties of the light kaonic nuclei are presented in Sec.~\ref{sec:result}.
A summary is given in Sec.~\ref{sec:summary}.

\section{Two-Body Interactions}
\label{sec:two-body}

\subsection{Hamiltonian and expectation values}

The Hamiltonian for $(\mathscr{N}-1)$ nucleons and an antikaon takes
 the form
\begin{align}
  H&=\sum_{i=1}^{\mathscr{N}}T_i-T_{\text{cm}}\notag\\
  &\quad +\sum_{i<
   j}^{\mathscr{N}-1}V_{ij}^{(NN)}+\sum_{i=1}^{\mathscr{N}-1}V_{i\mathscr{N}}^{(\bar{K}N)}
 +\sum_{i<
   j}^{\mathscr{N}}V^{\text{Coul.}}_{ij}.
\end{align}
Here $T_i$ is the kinetic energy of the $i$-th particle.
  The particle label, $i=\mathscr{N}$,
  always indicates an antikaon and
  the others are for nucleons.
$T_{\text{cm}}$ is the energy of the center-of-mass (c.m.) motion
\begin{align}
  T_{\text{cm}}=\frac{(\sum_{i=1}^{\mathscr{N}}{\bm p}_i )^2}{2\{(\mathscr{N}-1)m_N + m_{\bar{K}}\}},
\end{align}
where the isospin-averaged
nucleon and antikaon masses, $m_N=939$ MeV and $m_{\bar{K}}=496$ MeV, are used in this
paper.
$V_{ij}^{(NN)}$, $V_{ij}^{(\bar{K}N)}$ , and $V_{ij}^{\rm Coul.}$ are
the $NN$, $\bar{K}N$, and Coulomb 
 interactions
between the $i$- and $j$-th 
particles, respectively.
The $NN$ and $\bar{K}N$ interactions 
depend on isospin of two-particles,
and they can be written as
\begin{align}
 V_{ij} &= V^{I=0}_{ij}\hat{P}^{I=0}_{ij}+V^{I=1}_{ij}\hat{P}^{I=1}_{ij}\nonumber\\
 &=\frac{1}{2}(V^{I=0}_{ij}+V^{I=1}_{ij})-\frac{1}{2}(V^{I=0}_{ij}-V^{I=1}_{ij})\hat{P}_{\tau}^{ij},
\end{align}
where $\hat{P}^{I=0}_{ij}=\frac{1-\bm{\tau}_i\cdot\bm{\tau}_j}{4}$ and
$\hat{P}^{I=1}_{ij}=\frac{3+\bm{\tau}_i\cdot\bm{\tau}_j}{4}$ are isospin-projection
operators for $I=0$ and $1$, and
$\hat{P}_{\tau}^{ij}=\frac{1+\bm{\tau}_i\cdot\bm{\tau}_j}{2}$ is the isospin-exchange operator
for the $i$- and $j$-th particles.
The isospin-exchange operator $\hat{P}_{\tau}$ acts on particle basis as
$\hat{P}_{\tau}|nn\rangle=|nn\rangle$,
$\hat{P}_{\tau}|pp\rangle=|pp\rangle$
and $\hat{P}_{\tau}|pn\rangle=|np\rangle$ for $NN$,
and $\hat{P}_{\tau}|K^-n\rangle=|K^-n\rangle$, $\hat{P}_{\tau}|\bar{K}^0p\rangle=|\bar{K}^0p\rangle$,
$\hat{P}_{\tau}|K^-p\rangle=-|\bar{K}^0n\rangle$ and
$\hat{P}_{\tau}|\bar{K}^0n\rangle=-|K^-p\rangle$ for $\bar{K}N$. 
We have to treat the
$K^{-}p$-$\bar{K^{0}}n$ channel coupling explicitly in the
particle basis calculation.

The single-channel $\bar{K}N$ potential $V^{(\bar{K}N)}$ has an imaginary part which represents the decay processes into the lower energy $\pi\Sigma$ and $\pi\Lambda$ channels. Because of the complex nature of the potential, the Hamiltonian is non-Hermite and can have an eigenstate with a complex eigenvalue, called a quasi-bound state. In order to discuss the structure of the quasi-bound state, we need to evaluate the expectation values of some operators. For a stable bound state, the expectation value of an operator $\hat{\mathcal{O}}(\bm{x})$ with the wavefunction $\Psi_{JMM_T}(\bm{x})$ is given by (notation of the wavefunction will be explained in Sec.~\ref{sec:SVM})
\begin{align}
  \langle\hat{\mathcal{O}}\rangle&\equiv \int
 d{\bm x}[\Psi_{JMM_T}({\bm
 x})]^{*}\hat{\mathcal{O}}({\bm x})\Psi_{JMM_T}({\bm
 x}) ,\label{eq:standard_ex}
\end{align}
with the normalization of the wavefunction
\begin{align}
 1&=\int d{\bm x}|\Psi_{JMM_T}({\bm x})|^2.
\end{align}
However, since the eigenfunctions of a non-Hermite Hamiltonian do not form an orthogonal set~\cite{Hokkyo:1965}, we should introduce the Gamow states to treat an unstable state. The expectation value with the Gamow states is
\begin{align}
 \langle\hat{\mathcal{O}}\rangle_G&\equiv 
 \int d{\bm x}\Psi_{JMM_T,G}({\bm
 x})
 \hat{\mathcal{O}}({\bm x})\Psi_{JMM_T,G}({\bm
   x})
\end{align}
with the normalization
\begin{align}
1&=
 \int d{\bm x}[\Psi_{JMM_T,G}({\bm x})]^2.\label{Gamow}
\end{align}
With the normalization of Eq.~(\ref{Gamow}), expectation values are in general obtained as complex numbers, which are not straightforwardly interpreted. In some cases, however, we can extract a real-valued quantity. As explained in Appendix of Ref.~\cite{Miyahara:2015bya}, for a quasi-bound state whose real part of the eigenenergy is negative, the damping of the wavefunction outside the potential can be extracted from the standard expectation values with the normalization~\eqref{eq:standard_ex}. In this paper, we calculate the root-mean-square (rms) distances $\sqrt{\langle r^2\rangle}$, density distributions $\rho(r)$, and the probabilities of finding various channels in the wave functions $P$ by using the standard expectation value~(\ref{eq:standard_ex}). For the other operators such as Hamiltonian and its decomposition, the expectation values are calculated by using Gamow state normalization~(\ref{Gamow}).

\subsection{$NN$ interactions}

As a nucleon-nucleon interaction $V_{NN}$ we employ the Argonne 
V4'
potential~\cite{Wiringa:2002ja}.
AV4' potential is obtained by simplifying the full AV18 potential by
suppressing the small electro-magnetic, the spin-orbit, and the tensor terms and
readjusting the central spin- and isospin- dependent interactions.
TABLE~\ref{tab:nuclei} lists the binding energies and radii
  of two- to six-nucleon systems calculated with the AV4' potential.
The AV4' potential model reasonably reproduces the properties of light
nuclei.

\begin{table*}[htb]
\begin{center}
 \caption{Total binding energies $B$, point-proton rms radii
 $\sqrt{\langle r_p^2\rangle}$, 
 point-neutron rms radii $\sqrt{\langle
 r_n^2\rangle}$ and rms matter radii $\sqrt{\langle
 r_N^2\rangle}$
 of ordinary nuclei.
 The experimental data of the binding energies $B$
  are taken from Ref.~\cite{Audi:2002rp}.
  The nuclear charge radii $\sqrt{\langle r_{ch}^2\rangle}$
  in Ref.~\cite{Angeli201369} are converted into point-proton rms
 radii $\sqrt{\langle r_p^2\rangle}$ by using the formula:
 ${\langle r_{ch}^2\rangle} = {\langle r_p^2\rangle} +
 {\langle r_{pc}^2\rangle} + \frac{N}{Z}{\langle r_{nc}^2\rangle}+\frac{3\hbar}{4m_p^2c^2}$~\cite{PhysRevA.56.4579}, where $\langle r_{pc}^2\rangle=0.878^2$
 [fm$^2$], $\langle r_{nc}^2\rangle=-0.115$
 [fm$^2$], $\frac{3\hbar}{4m_p^2c^2}=0.0332$ [fm$^2$] are proton
 mean-square charge radius, neutron mean-square charge radius and
 Darwin-Foldy term, respectively.
 }
{\tabcolsep = 2.7mm
  \begin{tabular}{cccccccc} \hline\hline
    & AV4' & Expt. & AV4' & Expt. & \multicolumn{2}{c}{AV4'}    \\
   & \multicolumn{2}{c}{$B$ [MeV]} & \multicolumn{2}{c}{$\sqrt{\langle r_p^2\rangle}$ [fm]}
   & $\sqrt{\langle r_n^2\rangle}$ [fm] & $\sqrt{\langle r_N^2\rangle}$ [fm] 
    \\
   \hline
  $^2$H     & $ 2.24$ &  $ 2.22$ & $2.02$ & $1.97$ & $2.02$ & $2.02$   \\
  $^3$H     & $ 8.99$ &  $ 8.48$ & $1.59$ & $1.59$ & $1.70$ & $1.67$   \\
  $^3$He    & $ 8.33$ &  $ 7.72$ & $1.73$ & $1.77$ & $1.60$ & $1.69$   \\
  $^4$He    & $32.1 $ &  $28.3$ & $1.39$ & $1.46$ & $1.38$ & $1.39$   \\
  $^6$He    & $32.2 $ &  $29.3$ & $2.00$ & $1.93$ & $2.95$ & $2.67$   \\
  $^6$Li    & $35.8 $ &  $32.0$ & $2.43$ & $2.45$ & $2.42$ & $2.43$   \\
 \hline\hline
\end{tabular}  }
\label{tab:nuclei} 
\end{center}
\end{table*}

\subsection{$\bar{K}N$ interactions}\label{subsec:KNint}

As a realistic $\bar{K}N$ interaction, $V^{(\bar{K}N)}$, 
we employ the SIDDHARTA potential, which is the
energy-dependent effective interaction based on the chiral SU(3) dynamics
constructed in Ref.~\cite{Miyahara:2015bya}:
\begin{align}
 V^{(\bar{K}N)}(r,E) &=
 \frac{1}{\pi^{3/2}b^3}e^{-r^2/b^2}\frac{m_N}{2(E+m_N+m_{\bar{K}})}\nonumber\\
 &\times\frac{\omega_{\bar{K}}+E_N}{\omega_{\bar{K}}E_N}
 \left[\sum_i K_i \left(\frac{E}{100~\text{MeV}}\right) ^i  \right],
\end{align}
where $E$, $E_N$ and $\omega_{\bar{K}}$ are the non-relativistic two-body energy,
the energy of the nucleon and the energy of the antikaon:
\begin{align}
 E&=\sqrt{s}-m_N-m_{\bar{K}},\\
 E_N&=\frac{s-m_{\bar{K}}^2+m_N^2}{2\sqrt{s}},\\
 \omega_{\bar{K}}&=\frac{s-m_{N}^2+m_{\bar{K}}^2}{2\sqrt{s}}.
\end{align}
The coefficients $K_i$ of the energy dependent strength and the range parameter $b$ are determined so as to reproduce
the $\bar{K}N$ amplitude~\cite{Ikeda:2011pi,Ikeda:2012au} calculated 
based on the NLO chiral SU(3) dynamics (see
Ref.~\cite{Miyahara:2015bya}).
The SIDDHARTA potential is the single channel $\bar{K}N$
interaction model where the meson-baryon channel coupling effect with
strangeness $S=-1$ is renormalized, and thus the coefficients $K_i$ are the
complex numbers.
The origin of the energy dependence is two-fold. The coupled-channel interaction depends on the energy through the time derivatives in the chiral Lagrangians, and the construction of the equivalent single-channel potential introduces additional energy dependence.
By solving the Schr\"odinger equation, the 
pole positions of the $\bar{K}N(I=0)$ amplitude are found to be $1424-26i$
 and $1381-81i$ MeV.

For the use of the energy-dependent potential,
it is necessary to determine the $\bar{K}N$ two-body energies in the
$\mathscr{N}$-body systems.
Though
the two-body energies in the $\mathscr{N}$-body systems
cannot be determined uniquely,
we follow the same way as used in
Refs.~\cite{Dote:2008in,Dote:2008hw}
to determine
 the $\bar{K}N$ two-body energies
for practical calculations.\footnote{We also examine the prescription of the two-body energy suggested in Ref.~\cite{Barnea:2012qa}. The results of the few-body systems turn out to be in between the two choices shown in this paper.}
First, we introduce an ``antikaon binding energy'' $B_{\bar{K}}$ as
\begin{align}
 -B_{\bar{K}}\equiv \langle H\rangle_G - \langle H_N\rangle_G ,
\end{align}
where $H_N$ is the Hamiltonian for $(\mathscr{N}-1)$ nucleons defined by
\begin{align}
 H_N=\sum_{i=1}^{\mathscr{N}-1}T_i&+\sum_{i<
 j}^{\mathscr{N}-1}V_{ij}+
 V_{\text{Coulomb}}^{NN}-T_{\text{cm}}^N
 \end{align}
   with
\begin{align}
 T_{\text{cm}}^N&=\frac{(\sum_i^{\mathscr{N}-1}{\bm p}_i)^2}{2(\mathscr{N}-1)m_N}.
\end{align}
Note that $-B_{\bar{K}}$ is in general complex.
We employ the following three types of
the $\bar{K}N$ two-body energy as
 \begin{align}
  \sqrt{s}&=m_N+m_{\bar{K}}+\delta \sqrt{s}, \label{eq:sqrts}\\
 &\text{Type I:}~ \delta\sqrt{s}=-B_{\bar{K}},\\
 &\text{Type II:}~\delta\sqrt{s}=-B_{\bar{K}}/(\mathscr{N}-1).
 \end{align}
 Type I corresponds to the picture in which the $\bar{K}$ field collectively
 surrounds the $(\mathscr{N}-1)$ nucleons, and Type II corresponds to the picture in
 which the $\bar{K}$ energy is distributed equally
 to the $(\mathscr{N}-1)$ nucleons~\cite{Dote:2008in,Dote:2008hw}.
The eigenstate is determined in a self-consistent manner; the two-body energy calculated by the expectation values in Eq.~\eqref{eq:sqrts} should equal to the energy variable in the $\bar{K}N$ interaction in $V^{(\bar{K}N)}(r,E)$.

For comparison, we also examine the Akaishi-Yamazaki (AY) potential. The potential was originally constructed in the coupled-channel $\bar{K}N$-$\pi\Sigma$-$\pi\Lambda$ system by fitting the old data of the scattering lengths in Ref.~\cite{Martin:1980qe} and the nominal pole position of $\Lambda(1405)$~\cite{Akaishi:2002bg}. Here we adopt the single-channel version presented in Ref.~\cite{Yamazaki:2007cs} in which the energy dependence of the potential through the Feshbach projection method is eliminated by hand.

We note that the imaginary parts of the SIDDHARTA and AY potentials represent the decay into the $\pi\Sigma$ and $\pi\Lambda$ channels. 
In the few-body kaonic nuclei, there are two types of the decay processes, the mesonic decays with a pion emission and the nonmesonic decays with multi-nucleon absorptions. In this work, the imaginary part of the eigenenergy corresponds only to the mesonic decay width, reflecting the imaginary part of the two-body potential.
When the nonmesonic decays are taken into account, such effect
would increase the decay width of the kaonic nuclei by several tens of MeV~\cite{Dote:2008hw,Sekihara:2009yk,Sekihara:2012wj,Bayar:2012rk}.

\section{Stochastic variational method with correlated Gaussian basis}
\label{sec:SVM}
We investigate the structure of the kaonic nuclei 
with a powerful few-body approach, that is, the 
SVM with the CG~\cite{Varga:1995dm,Varga:1997xga,suzuki1998stochastic}.
The method is flexible to cope with strongly correlated
few-particle quantum systems as exemplified in Ref.~\cite{RevModPhys.85.693}.

The wavefunction for the $\mathscr{N}$-body system is expanded as a combination
of the basis functions:
\begin{align}
 \Psi_{JMM_T}({\bm x})=\sum^K_{k=1}c_k\Phi_{JMM_T}({\bm x},A_k) ,
\end{align}
where $J$ is the total angular momentum, $M$ ($M_{T}$) is the $z$-component of the total angular momentum (isospin).
Since
we employ central $NN$ and $\bar{K}N$ interactions,
 no channel coupling occurs between states with different $L$.
 In this paper, we consider total orbital momentum $L=0$ state by
  taking the basis functions with total spin $J(=S)$ 
to have the form
\begin{align}
  \Phi_{SM_SM_T}({\bm x},A)&=\mathscr{A}\{\exp (-\widetilde{{\bm x}} A{\bm x})
  \chi_{SM_S}\eta_{M_T}\},\label{eq:basis}
\end{align}
where the operator $\mathscr{A}$ is an antisymmetrizer for the 
nucleons;
$M_S(=M)$
is the $z$-components of the total spin; 
${\bm x}$ is an $(\mathscr{N}-1)$-dimensional column vector,
whose $i$-th element is a $3$-dimensional Jacobi coordinate ${\bm x}_i$;
the symbol $\widetilde{\bm x}$ stands for a transpose of 
${\bm x}$;
$A$ is
an $(\mathscr{N}-1)\times(\mathscr{N}-1)$-positive-definite-symmetric matrix.
The Jacobi coordinate, ${\bm x}_i$, including
 the center-of-mass coordinate ${\bm x}_{\mathscr{N}}$
are related to the $i$-th single-particle
coordinate ${\bm r}_i$ by a linear transformation:
\begin{align}
  {\bm x}_i&=\sum_{j=1}^{\mathscr{N}}U_{ij}{\bm r}_j
\end{align}
with
\begin{widetext}
\begin{align}
 U&=
   \begin{pmatrix}
  1& -1 & 0& \cdots&0\\
  \frac{1}{2}&\frac{1}{2}&-1&\cdots&0\\
  \vdots&  &\ddots &\ddots &\vdots\\
  \frac{1}{\mathscr{N}-1}&\cdots &\cdots &\frac{1}{\mathscr{N}-1} &-1 \\
  \frac{m_N}{(\mathscr{N}-1)m_N+m_{\bar{K}}}&\cdots &\cdots &\frac{m_N}{(\mathscr{N}-1)m_N+m_{\bar{K}}} &\frac{m_{\bar{K}}}{(\mathscr{N}-1)m_N+m_{\bar{K}}}
\end{pmatrix}~~.
\end{align}
\end{widetext}
We can easily apply the CG to the present three- to seven-body
  model because the CG keeps
  its functional form under any linear transformation
  between different coordinate sets
  for any number of particles.
The CG basis in Eq.~(\ref{eq:basis}) can only be applicable to states
with total orbital angular momentum $L=0$.
It should be noted that 
the higher partial waves for each coordinate are
taken into account through the cross terms, ${\bm x}_i\cdot {\bm x}_j$.

The spin wavefunction $\chi_{SM_S}$
is expressed using the basis of successive coupling:
\begin{align}
 \chi_{SM_S}=\big|\big[\cdots\big[\big[\tfrac{1}{2}\tfrac{1}{2}\big]_{S_{12}}
 \tfrac{1}{2}\big]_{S_{123}}\cdots\big]_{SM_S}\big\rangle.
\end{align}
Here we take all possible 
intermediate spins $(S_{12},~S_{123},~\dots)$ 
 for a given $S$.
For the isospin wavefunction, $\eta_{M_T}$,
we employ the particle basis, which
is given as the product of single-particle isospin wavefunctions:
\begin{align}
 \eta_{M_T}=\eta_{\frac{1}{2}m_{\tau_1}}\cdots\eta_{\frac{1}{2}m_{\tau_{\mathscr{N}}}}.
\end{align}
The sets of the
single-particle isospins 
$(m_{\tau_1},~\dots,~m_{\tau_{\mathscr{N}}})$ take the
values
\begin{align}
 m_{\tau_k}=
   \begin{cases}
    \frac{1}{2}  & (k=1,~\dots,~M_T+\frac{\mathscr{N}}{2}-1,~\mathscr{N})\\
    -\frac{1}{2} & (\text{otherwise}),
   \end{cases}\label{isospin_k0}
\end{align}
for the states with $\bar{K}^{0}$ and
\begin{align}
 m_{\tau_k}=
   \begin{cases}
    \frac{1}{2}  & (k=1,~\dots,~M_T+\frac{\mathscr{N}}{2})\\
    -\frac{1}{2} & (\text{otherwise}),
   \end{cases}\label{isospin_k-}
\end{align}
for the states with $K^{-}$.

 \begin{figure}[tbh]
\includegraphics[width=0.5\textwidth,clip]{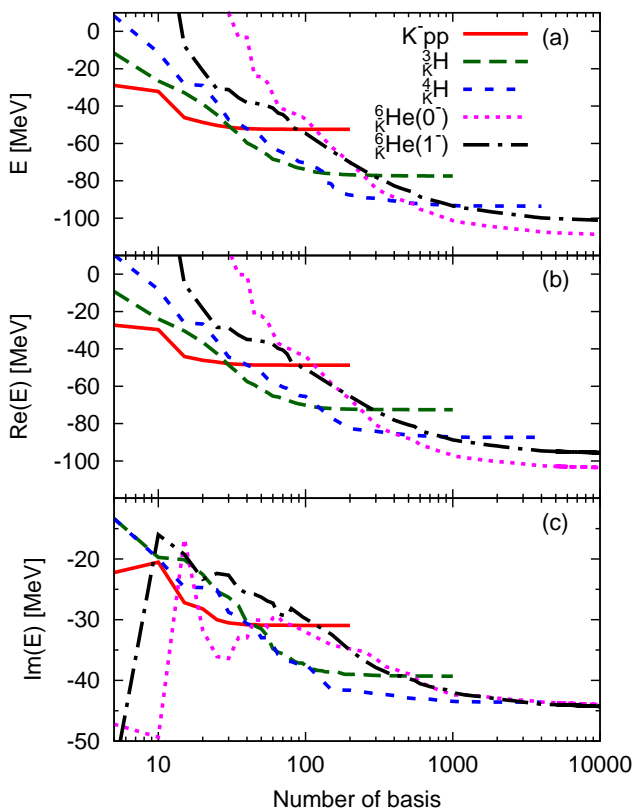}
\caption{(Color online) Energy convergence of the ground states
  of three- to seven-body systems 
  including an antikaon.
  (a) eigenvalues calculated only with the real part of the Hamiltonian;
  (b) real part of 
  the full Hamiltonian eigenvalue; (c) imaginary
  part of the full Hamiltonian eigenvalue.
The AY potential model is employed
as the $\bar{K}N$ interaction.
See text for details.
  }
 \label{fig:ener_conv}
 \end{figure}

Each basis function has $\mathscr{N}(\mathscr{N}-1)/2$
nonlinear parameters $(A_k)_{ij}$ and also spin and isospin quantum numbers.
The adequate choice of these parameters is crucial to determine
accuracy of the variational calculation.
The SVM offers efficient and economical ways to find optimal sets
of the variational parameters~\cite{Varga:1995dm,Varga:1997xga,suzuki1998stochastic},
in which we increase the basis size one-by-one by searching for the best
among many random trials for the basis function.
For the Hermitian Hamiltonian, the eigenvalues for the trial
wavefunctions are larger than or equal to the 
exact eigenvalue.
Since we use the complex $\bar{K}N$ potential in this work,
  the eigenvalues of the trial wavefunctions give no longer
  the lower limit.
Practically,
we apply the SVM for the real part of the Hamiltonian to obtain the energy curve,
and then we diagonalize the full Hamiltonian by using the basis 
optimized for the ground state with
the real Hamiltonian.
The validity of this method can be confirmed in the two-body sector where the exact value of the pole position can be obtained.
Examples of the few-body calculations for
$\bar{K}NN$ ($K^-pp$), $\bar{K}NNN$ ($^3_{\bar{K}}$H), $\bar{K}NNNN$ ($^4_{\bar{K}}$H) and $\bar{K}NNNNNN$ [$^6_{\bar{K}}$He ($J^\pi$)]
are shown in Fig.~\ref{fig:ener_conv} with the AY potential. The eigenvalues with the real part of the Hamiltonian 
are shown in Fig.~\ref{fig:ener_conv} (a).
Corresponding complex energy curves of the full Hamiltonian
  are plotted in Figs.~\ref{fig:ener_conv} (b) and (c).
We find that if the energy convergence is reached
  with the real part of the Hamiltonian,
the eigenvalues with the full Hamiltonian are
also converged.
The obtained energies in this method are
consistent with other
calculations for two- and three-body
systems~\cite{Akaishi:2002bg,Yamazaki:2002uh,Yamazaki:2007cs}.

For the $\bar{K}NN$, $\bar{K}NNN$, $\bar{K}NNNN$  and $\bar{K}NNNNNN$
systems,
the basis sizes are $200$, $1000$, $4000$ and $10000$, respectively.
The binding energies and widths change less than $0.0001$ MeV when
the numbers of basis increase by one from these basis numbers.

\section{Structure of few-body systems}
\label{sec:structure}
The internal structure of the the $\bar{K}$ nuclei is reflected in the obtained wavefunction $\Psi_{JMM_T}$. Here we define several quantities which are useful to investigate the structure of the few-body systems.

We first define the $NN$ root-mean-square (rms) distances $\sqrt{\langle r_{NN}^2 \rangle}$,
$\bar{K}N$ rms distances $\sqrt{\langle r_{\bar{K}N}^2 \rangle}$, $N$ rms
radii $\sqrt{\langle r_{N}^2 \rangle}$ and $\bar{K}$ rms radii $\sqrt{\langle r_{\bar{K}}^2 \rangle}$ by using the following operators:
\begin{align}
 r_{NN}^2&=\sum_{i<j}^{\mathscr{N}-1}\frac{2|{\bm r}_i-{\bm  r}_j|^2}{(\mathscr{N}-1)(\mathscr{N}-2)},\\
 r_{\bar{K}N}^2&=\sum_{i}^{\mathscr{N}-1}\frac{|{\bm r}_{\bar{K}}-{\bm
 r}_i|^2}{(\mathscr{N}-1)},\\ 
 r_{N}^2&=\sum_{i}^{\mathscr{N}-1}\frac{|{\bm r}_i-{\bm  x}_\mathscr{N}|^2}{(\mathscr{N}-1)},\\
 r_{\bar{K}}^2&= |{\bm r}_{\bar{K}}-{\bm x}_\mathscr{N}|^2,
\end{align}
where ${\bm r}_{i}$ and ${\bm r}_{\bar{K}}={\bm r}_\mathscr{N}$ are the single-particle coordinates of the $i$-th nucleon and the antikaon.
The rms distances represent the averaged distance of the two-body subsystems, and the rms radii measure the averaged distance of the particle from the center-of-mass of the total system $\bm{x}_{\mathscr{N}}$. As discussed in Sec.~\ref{sec:two-body}, we calculate the expectation values of these operators $\sqrt{\langle r^2\rangle}$ using the standard normalization condition~(\ref{eq:standard_ex}).

To investigate how the nuclear system shrinks by adding an antikaon,
we define the nucleon density distributions
\begin{align}
  \rho_N^{N\text{cm}}(r)&=\sum_{i=1}^{\mathscr{N}-1}\int d{\bm x}|\Psi_{JMM_T}({\bm
  x})|^2\delta({\bm r}_{i}^{N\text{cm}}-{\bm r}),
  \label{eq:rhoNcm} \\
  {\bm r}_{i}^{N\text{cm}}&={\bm r}_{i}-({\bm x}_{\mathscr{N}-1}+{\bm r}_{\bar{K}}),
\end{align}
where ${\bm r}_{i}^{N\text{cm}}$ denotes the $i$-th nucleon coordinate measured
from the center-of-mass system of nucleons;
$\rho_N^{N\text{cm}}(r)$ is normalized as
$\int4\pi r^2\rho_N^{N\text{cm}}(r)=\mathscr{N}-1$. 
Here, again, we adopt the standard normalization condition~(\ref{eq:standard_ex}). The comparison of $\rho_N^{N\text{cm}}(r)$ with the corresponding quantity of the normal nuclei with $\mathscr{N}-1$ nucleons shows the effect of the modification of the distribution of the nucleons by the presence of the antikaon.

We also calculate the nucleon and antikaon density distribution $\rho_N$ and
 $\rho_{\bar{K}}$ measured from the total center-of-mass system defined as:
\begin{align}
  \rho_N(r)&=\sum_{i=1}^{\mathscr{N}-1}\int d{\bm x}|\Psi_{JMM_T}({\bm
  x})|^2\delta({\bm r}_{i}^{\text{cm}}-{\bm r}),\label{eq:rhoN}\\
  \rho_{\bar{K}}(r)&=\int d{\bm x}|\Psi_{JMM_T}({\bm
  x})|^2\delta({\bm r}_{\bar{K}}^{\text{cm}}-{\bm r}),
  \label{eq:rhoK}
\end{align}
where ${\bm r}_{i(\bar{K})}^{\text{cm}}={\bm r}_{i(\bar{K})}-{\bm x}_{\mathscr{N}}$ is
$i$-th nucleon (antikaon) coordinate from the total center-of-mass coordinate ${\bm x}_{\mathscr{N}}$.

It is also instructive to estimate the fractions of different components in the wavefunctions. We define the projections onto the component with $K^{-}$ [Eq.~(\ref{isospin_k-})] and that with $\bar{K}^{0}$ [Eq.~\eqref{isospin_k0}] as
\begin{align}
   \hat{P}_{K^{-}}
   &=\frac{1}{2}(1-\tau_{\mathcal{N}}^{(3)}), \\
   \hat{P}_{\bar{K}^{0}}
   &=\frac{1}{2}(1+\tau_{\mathcal{N}}^{(3)}),
\end{align}
with $\hat{P}_{K^{-}}+\hat{P}_{\bar{K}^{0}}=1$. 
By taking the expectation value of Eq.~(\ref{eq:standard_ex}), we obtain the probability of finding each component in the wavefunction 
\begin{align}
   P_{K^{-}}
   &=\langle\hat{P}_{K^{-}}\rangle \\
   P_{\bar{K}^{0}}
   &=\langle\hat{P}_{\bar{K}^{0}}\rangle .
\end{align}
The projection operators can also be used to decompose the eigenenergy into different contributions from each term of the Hamiltonian. For this purpose, we use the Gamow state normalization~(\ref{Gamow}). The expectation values of the kinetic energy and potential
energy of the diagonal $K^-$ channel $\langle T\rangle^{K^-}_G$ and $\langle
V\rangle^{K^-}_G$, and of the diagonal $\bar{K}^0$ channel
$\langle T\rangle_G^{\bar{K}^0}$ and  $\langle
V\rangle_G^{\bar{K}^0}$,
and of the off-diagonal $K^-$-$\bar{K}^0$ channel $\langle
V\rangle_G^{K^-\bar{K}^0}$ are given by
\begin{align}
   &\begin{pmatrix}
    \langle T + V\rangle^{K^-}_G
    & \langle V\rangle_G^{K^-\bar{K}^0}\\
    \langle V\rangle_G^{K^-\bar{K}^0}
    & \langle T + V\rangle_G^{\bar{K}^0}
   \end{pmatrix}\nonumber\\
   &\equiv
   \begin{pmatrix}
    \langle \hat{P}_{K^-}(T + V)\hat{P}_{K^-} \rangle_G
    & \langle \hat{P}_{K^-}V\hat{P}_{\bar{K}^0}\rangle_G\\
    \langle\hat{P}_{\bar{K}^0} V\hat{P}_{K^-}\rangle_G
    & \langle \hat{P}_{\bar{K}^0}(T + V)\hat{P}_{\bar{K}^0}\rangle_G
   \end{pmatrix}.
  \end{align}
With these definitions, the eigenenergy is decomposed as 
\begin{align}
   -B-i\frac{\Gamma}{2}
   &=\langle T\rangle^{K^-}_G 
   +\langle T\rangle^{\bar{K}^0}_G
   +\langle V\rangle^{K^-}_G
   +\langle V\rangle^{\bar{K}^0}_G
   +2\langle V\rangle_G^{K^-\bar{K}^0} .
   \label{eq:decomposition}
\end{align}
We also investigate the probability of finding each $\bar{K}N$ isospin
component in the wavefunction by using the following expectation values,
\begin{align}
   P_{\bar{K}N}^{I=0}
   &=\sum_i^{\mathcal{N}-1}\frac{\langle \hat{P}^{I=0}_{i\mathcal{N}}\rangle}{\mathcal{N}-1}, \\
   P_{\bar{K}N}^{I=1}
   &=\sum_i^{\mathcal{N}-1}\frac{\langle \hat{P}^{I=1}_{i\mathcal{N}}\rangle}{\mathcal{N}-1}.
\end{align}
\section{Results and Discussion}
\label{sec:result}
\subsection{Structure of strange dibaryon resonances $\bar{K}NN$}
\label{subsec:KNN}
\begin{table*}[tb]
\begin{center}
\caption{Properties of the calculation for $K^-pp$-$\bar{K}^0pn$ system
 with $J^\pi=0^-$.
 See text for details.}
{\tabcolsep = 2.7mm
  \begin{tabular}{cccccccccc} \hline\hline
   \multicolumn{4}{c}{$K^-pp$-$\bar{K}^0pn~(J^\pi=0^-)$} \\\hline
   Model & \multicolumn{2}{c}{SIDDHARTA}&AY\\
    & Type I  & Type II &  \\
   \hline
  $B$ [MeV]     &$27.9$ & $26.1$  &$48.7$    \\
  $\Gamma$ [MeV] &$30.9$ & $59.3$ &$61.9$       \\
  $\delta \sqrt{s}$ [MeV]     &$-61.0-i25.0$ & $-30.2-i23.7$   &   \\
  $P_{K^-}$ &$0.65$ & $0.65$  &$0.64$      \\
  $P_{\bar{K}^0}$ &$0.35$ & $0.35$  &$0.36$      \\
  $\sqrt{\langle r_{NN}^2\rangle}$ [fm] &$2.16$ & $2.07$  &$1.84$ \\
  $\sqrt{\langle r_{\bar{K}N}^2\rangle}$ [fm] &$1.80$ & $1.73$  &$1.55$  \\
  $\sqrt{\langle r_N^{2}\rangle}$ [fm] &$1.12$ & $1.08$  &$0.958$     \\
  $\sqrt{\langle r_{\bar{K}}^{2}\rangle}$ [fm] &$1.14$ & $1.10$  &$0.988$    \\
  $\langle T\rangle^{K^-}_G$ [MeV] &$117+i28.8$ & $124+i53.1$  &$102+i31.4$    \\
  $\langle V\rangle^{K^-}_G$ [MeV] &$-113-i33.7$ & $-120-i63.9$  &$-102-i47.0$    \\
  $\langle T\rangle_G^{\bar{K}^0}$ [MeV] &$74.3+i18.4$ & $76.3+i33.1$  &$63.1+i15.5$    \\
  $\langle V\rangle_G^{\bar{K}^0}$ [MeV] &$-62.0-i19.1$ & $-64.3-i35.6$  &$-48.6-i21.6$    \\
  $2\langle V\rangle_G^{K^-\bar{K}^0}$ [MeV] &$-44.1-i9.76$ & $-41.9-i16.4$  &$-64.0-i9.24$    \\
  $P_{\bar{K}N}^{I=0}$ &$0.72$ & $0.73$  &$0.73$    \\
  $P_{\bar{K}N}^{I=1}$ &$0.28$ & $0.27$  &$0.27$    \\
 \hline\hline
\end{tabular}  }
\label{tab:Kpp} 
\end{center}
\end{table*}
\begin{table*}[tb]
\begin{center}
\caption{Properties of the calculation for $K^-pn$-$\bar{K}^0nn$ system
 with $J^\pi=0^-$.}
{\tabcolsep = 2.7mm
  \begin{tabular}{cccccccccc} \hline\hline
   \multicolumn{4}{c}{$K^-pn$-$\bar{K}^0nn~(J^\pi=0^-)$} \\\hline
   Model& \multicolumn{2}{c}{SIDDHARTA}&AY\\
   & Type I  & Type II &  \\\hline
  $B$ [MeV]     &$27.6$ & $25.3$  &$48.1$    \\
  $\Gamma$ [MeV] &$31.6$ & $59.4$ &$61.6$       \\
  $\delta \sqrt{s}$ [MeV]     &$-60.2-i25.6$ & $-29.4-i23.8$   &  \\
  $P_{K^-}$ &$0.38$ & $0.38$  &$0.37$      \\
  $P_{\bar{K}^0}$ &$0.62$ & $0.62$  &$0.63$      \\
  $\sqrt{\langle r_{NN}^2\rangle}$ [fm] &$2.18$ & $2.10$  &$1.85$  \\
  $\sqrt{\langle r_{\bar{K}N}^2\rangle}$ [fm] &$1.82$ & $1.75$  &$1.56$  \\
  $\sqrt{\langle r_N^{2}\rangle}$ [fm] &$1.13$ & $1.09$  &$0.963$     \\
  $\sqrt{\langle r_{\bar{K}}^{2}\rangle}$ [fm] &$1.15$ & $1.11$  &$0.993$    \\
  $\langle T\rangle^{K^-}_G$ [MeV] &$67.6+i20.6$ & $81.4+i34.9$  &$65.3+i16.1$    \\
  $\langle V\rangle^{K^-}_G$ [MeV] &$-44.5-i10.2$ & $-69.9-i38.0$  &$-51.3-i22.6$    \\
  $\langle T\rangle_G^{\bar{K}^0}$ [MeV] &$112+i28.7$ & $118+i52.2$  &$99.5+i30.8$    \\
  $\langle V\rangle_G^{\bar{K}^0}$ [MeV] &$-107-i33.3$ & $-112-i62.2$  &$-97.3-i45.8$    \\
  $2\langle V\rangle_G^{K^-\bar{K}^0}$ [MeV] &$-44.5-i10.2$ & $-42.2-i16.7$  &$-64.3-i9.40$    \\
  $P_{\bar{K}N}^{I=0}$  &$0.72$ & $0.72$  &$0.73$    \\
  $P_{\bar{K}N}^{I=1}$  &$0.28$ & $0.28$  &$0.27$    \\
 \hline\hline
\end{tabular}  }
\label{tab:Kpn} 
\end{center}
\end{table*}
\begin{figure}[tbh]
 \includegraphics[width=0.48\textwidth,clip]{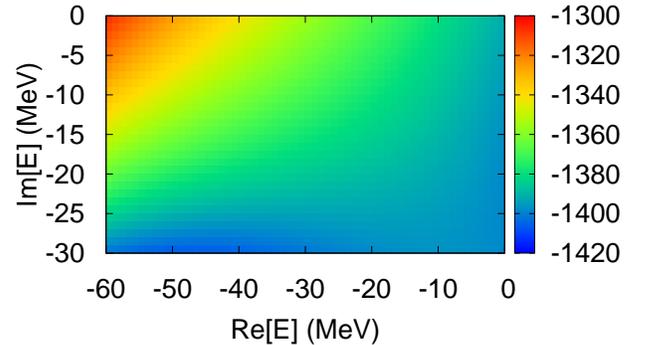}
 \caption{(Color online)
   Real part of the SIDDHARTA potential, ${\rm Re} V_{\bar{K}N}^{I=0}(r=0,E)$,
   on the complex energy plane.}
 \label{fig:MH_pot_re}
\end{figure}
\begin{figure*}[tbh]
 \includegraphics[width=0.48\textwidth,clip]{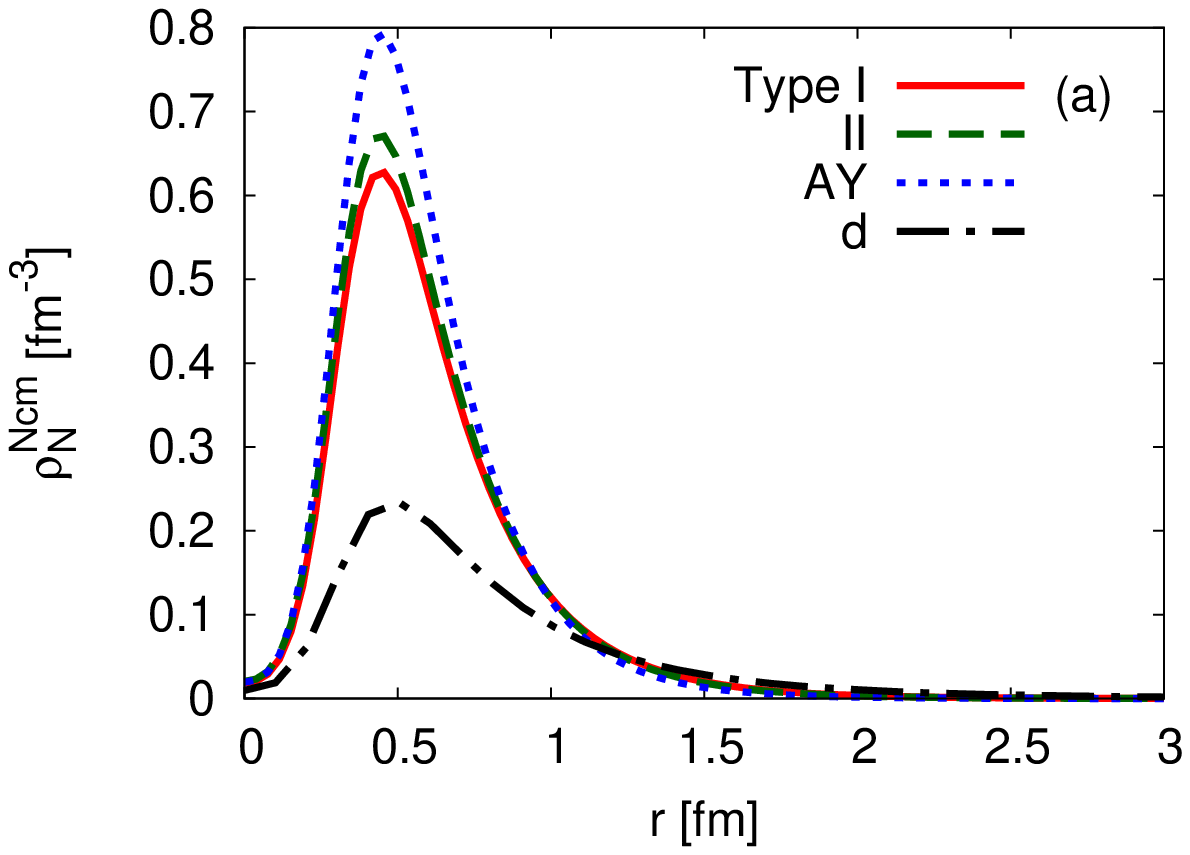}
 \includegraphics[width=0.48\textwidth,clip]{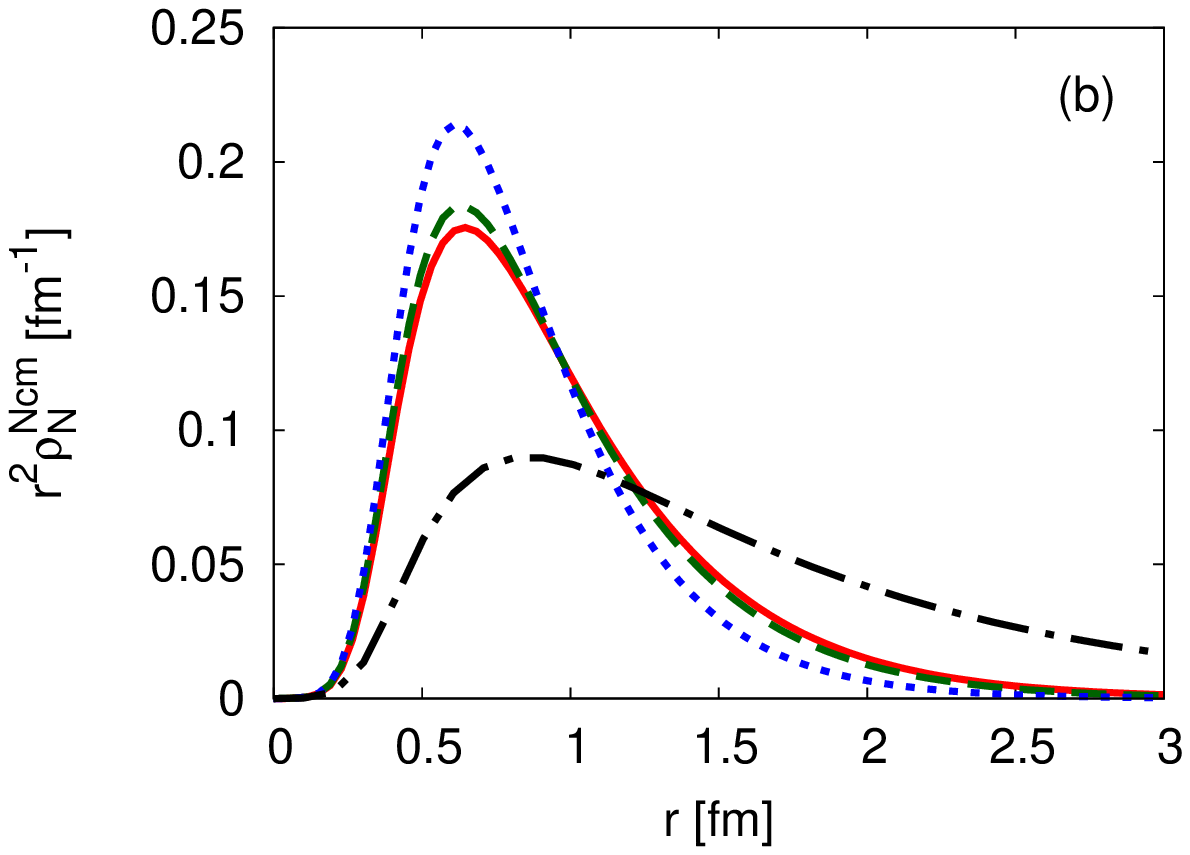}
 \caption{(Color online) Nucleon density distributions (a)
 $\rho_N^{N\text{cm}}(r)$ and (b) $r^2\rho_N^{N\text{cm}}(r)$ for
   $K^-pp$-$\bar{K}^0pn$ system measured from
   the center-of-mass of nucleons.
    We employ the SIDDHARTA potential with the Types I and II
    $\bar{K}N$ two-body energies and the AY potential.  
The dotted-dashed curves show that of deuteron for comparison.
 Note that the spin parity is  $J^\pi =0^-$ for $\bar{K}NN$ but
 $J^\pi=1^+$ for deuteron.}
 \label{fig:K-pp_N_dist}
\end{figure*}
\begin{figure*}[tbh]
 \includegraphics[width=0.48\textwidth,clip]{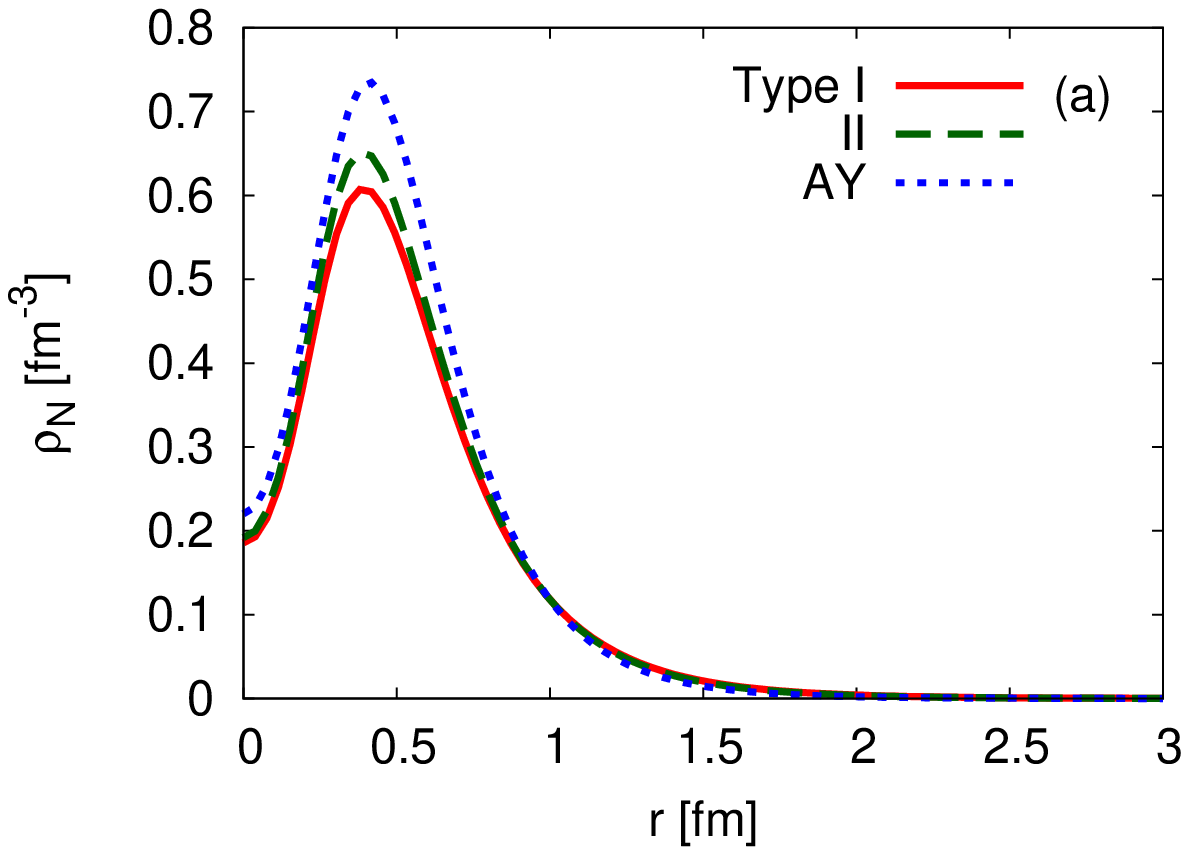}
 \includegraphics[width=0.48\textwidth,clip]{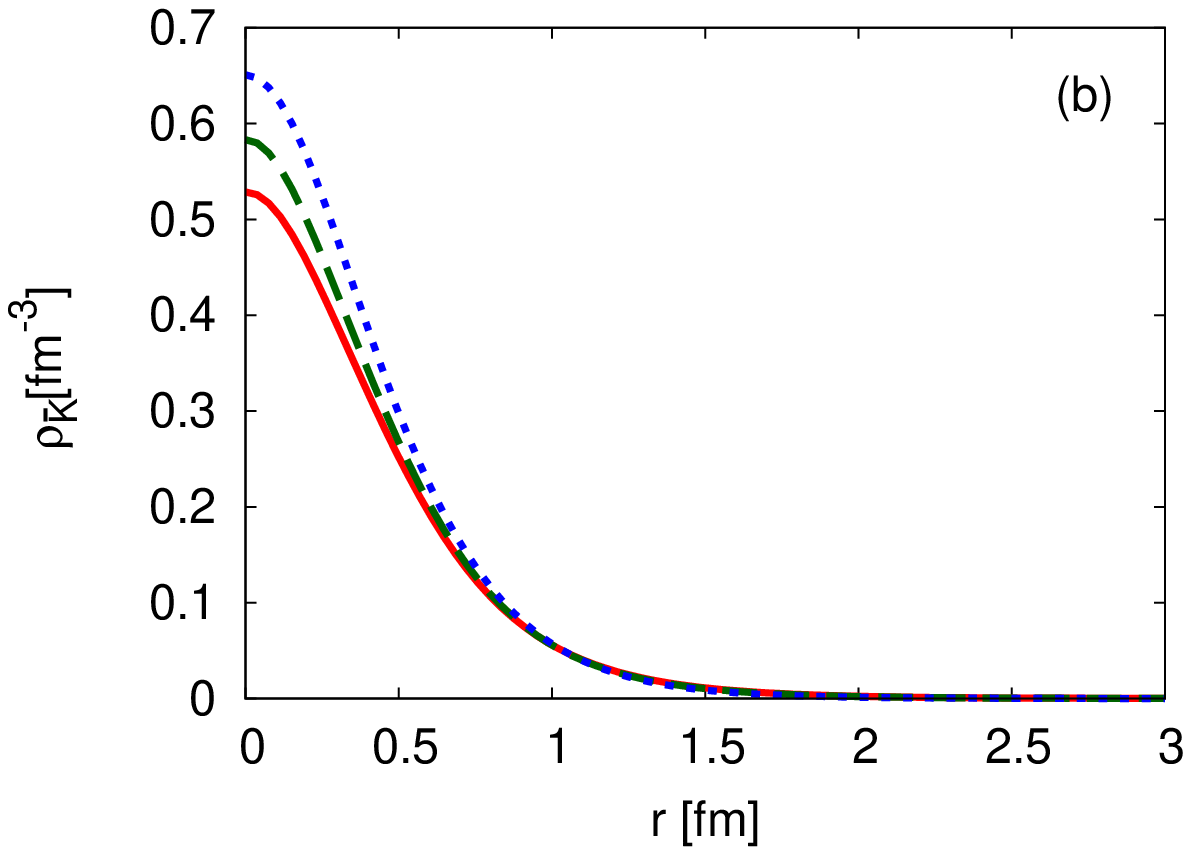}
 \caption{(Color online) (a) Nucleon density distribution
 $\rho_N(r)$ and (b) antikaon density distribution $\rho_{\bar{K}}(r)$ for
   $K^-pp$-$\bar{K}^0pn$ system measured from the center-of-mass
   of the system. The SIDDHARTA and AY potentials are employed.
 }
 \label{fig:K-pp_dist}
\end{figure*}

We proceed now to investigate the structure of the kaonic nuclei.
For the three-body systems,
we investigate the structure of the $I=1/2$ quasi-bound states
for the strange dibaryon resonances $\bar{K}NN$, $K^-pp$-$\bar{K}^0pn$ and $K^-pn$-$\bar{K}^0nn$ systems, with
$J^\pi = 0^-$ and $1^-$
in the charge-basis representation.
As in the previous studies~\cite{Shevchenko:2006xy,Shevchenko:2007ke,Ikeda:2007nz,Ikeda:2008ub,Yamazaki:2007cs,Dote:2008in,Dote:2008hw,Wycech:2008wf,Ikeda:2010tk,Barnea:2012qa,Dote:2014via},
we find one quasi-bound state below the
$\Lambda(1405)+N$ threshold for $J^\pi=0^-$,
but we could not find any states below the threshold for $J^\pi=1^-$.
We summarize the detailed 
properties of the 
$K^-pp$-$\bar{K}^0pn$ and
$K^-pn$-$\bar{K}^0nn$ systems with $J^\pi=0^-$ in Tables~\ref{tab:Kpp} and \ref{tab:Kpn}, respectively.

We first compare the results with different choices of the two-body energy
(Types I and II) discussed in Sec.~\ref{subsec:KNint}.
The real part 
of $\delta\sqrt{s}$ with Type II
is about a half of that of Type I. 
The binding energy of $K^-pp$-$\bar{K}^0pn$ system is not sensitive to
the choice of the two-body energy 
and the values are around $27$ MeV.
Meanwhile, the decay width of Type I
($\sim 31$ MeV) becomes about
a half of that of 
Type II ($\sim 59$ MeV).
The rms distances
$\sqrt{\langle r_{NN}^2 \rangle}$, $\sqrt{\langle r_{\bar{K}N}^2 \rangle}$,
$\sqrt{\langle r_{N}^2 \rangle}$, and
$\sqrt{\langle r_{K}^2 \rangle}$ of 
Type II are slightly smaller than
those of 
Type I. 
The probabilities of finding several components, $P_{K^-}$, $P_{\bar{K}^0}$, $P_{\bar{K}N}^{I=0}$ and $P_{\bar{K}N}^{I=1}$ 
are not sensitive to the choice of the two-body energy. 

We obtain the binding energies at 20-30 MeV, which are consistent with recent experimental measurement of the ${}^{3}{\rm He}(K^{-},\Lambda p)n$ reaction by J-PARC E15~\cite{Sada:2016nkb} and its theoretical analysis~\cite{Sekihara:2016vyd}.
The obtained binding energies with the
SIDDHARTA potential are $5$-$10$ MeV
larger than the values obtained in
Refs.~\cite{Dote:2008hw,Dote:2008in,Barnea:2012qa} with the
chiral potential 
constructed in Ref.~\cite{Hyodo:2007jq}.
This difference mainly comes from the 
different treatment of the imaginary part of the potential
as well as the difference of the potential model.
In Refs.~\cite{Dote:2008hw,Dote:2008in,Barnea:2012qa},
the Schr\"{o}dinger equation is solved only with the real part of
Hamiltonian and estimate the decay width
by taking the expectation value with the imaginary part of
  the potential,
  while we solve the Schr\"{o}dinger equation by direct diagonalization
  with the full complex Hamiltonian.
  If we do not take into account the energy dependence of the $\bar{K}N$ potential,
the binding energies obtained by direct diagonalization of the complex
potential $V_{\bar{K}N}(r,E)$ become
smaller than those obtained by using the real part of the potential $\text{Re}[V_{\bar{K}N}(r,E)]$.
However, the $\bar{K}N$ two-body energy $E$ in the three-body system
 also becomes a complex value with the complex potential $V_{\bar{K}N}(r,E)$.
Considering the energy dependence of the potential,
the $\bar{K}N$ interaction $V_{\bar{K}N}(r,E)$ becomes more attractive than
the $\bar{K}N$ interaction $V_{\bar{K}N}(r,\text{Re}[E])$ on the real energy axis.
This is found in Fig.~\ref{fig:MH_pot_re}
which plots the real part of $V_{\bar{K}N}$ on the complex energy plane.
As a result of self-consistent calculation, the binding
energies of the complex potential $V_{\bar{K}N}(r,E)$ become larger than those of only the real
part of potential $\text{Re}[V_{\bar{K}N}(r,E)]$.
Therefore, we obtain the binding energies $5$-$10$ MeV
larger than those obtained in
Refs.~\cite{Dote:2008hw,Dote:2008in,Barnea:2012qa}.

We then investigate the origin of the binding using the decomposition in Eq.~\eqref{eq:decomposition}. From Tables~\ref{tab:Kpp} and \ref{tab:Kpn}, we see that Re $\langle T\rangle^{K^-,\bar{K}^{0}}_G$ almost cancels out Re $\langle V\rangle^{K^-,\bar{K}^{0}}_G$ 
in both $K^{-}$ and $\bar{K}^{0}$ channels.
Therefore the $K^-$-$\bar{K}^0$ channel coupling is essential for
the energy gain.
Meanwhile, both of the diagonal and the off-diagonal components
contribute to 
the decay width.

It is also instructive to decompose the wavefunction into the isospin components.
The dominant isospin component of the 
$\bar{K}NN$ ground state is considered to have
$I_{NN}=1$ and total isospin $I=1/2$.
This is because the $I_{\bar{K}N}=0$ channel has stronger attraction than
that of the $I_{\bar{K}N}=1$ channel,
and the $I_{NN}=1$ channel gives more $I_{\bar{K}N}=0$ component
than that of the $I_{NN}=0$ channel~\cite{Yamazaki:2002uh}. This can easily be explained by
  recoupling the isospins in the following way:
\begin{align}
  [[\eta(N)_{\frac{1}{2}}&\eta(N)_{\frac{1}{2}}]_1\eta(\bar{K})_{\frac{1}{2}}]_{\frac{1}{2}} \nonumber\\
 &=\frac{\sqrt{3}}{2}[\eta(N)_{\frac{1}{2}}[\eta(N)_{\frac{1}{2}}\eta(\bar{K})_{\frac{1}{2}}]_0]_{\frac{1}{2}}
 \nonumber\\
 &+\frac{1}{2}[\eta(N)_{\frac{1}{2}}[\eta(N)_{\frac{1}{2}}\eta(\bar{K})_{\frac{1}{2}}]_1]_{\frac{1}{2}},
\end{align}
\begin{align}
  [[\eta(N)_{\frac{1}{2}}&\eta(N)_{\frac{1}{2}}]_0\eta(\bar{K})_{\frac{1}{2}}]_{\frac{1}{2}} \nonumber\\
 &=-\frac{1}{2}[\eta(N)_{\frac{1}{2}}[\eta(N)_{\frac{1}{2}}\eta(\bar{K})_{\frac{1}{2}}]_0]_{\frac{1}{2}}
 \nonumber\\
 &+\frac{\sqrt{3}}{2}[\eta(N)_{\frac{1}{2}}[\eta(N)_{\frac{1}{2}}\eta(\bar{K})_{\frac{1}{2}}]_1]_{\frac{1}{2}}.
\end{align}
If the ground state is a pure $I=1$ $NN$ state, the ratio of the probabilities of finding the $\bar{K}N$ $I=0$ and $I=1$ channels is given by $P_{\bar{K}N}^{I=0}:P_{\bar{K}N}^{I=1}= 3:1$. We can further decompose this state by the third component of the isospin of the antikaon as
\begin{align}
 [[\eta(N)_{\frac{1}{2}}&\eta(N)_{\frac{1}{2}}]_1\eta(\bar{K})_{\frac{1}{2}}]_{\frac{1}{2},\frac{1}{2}} \nonumber\\
 &=-\sqrt{\frac{2}{3}}[\eta(N)_{\frac{1}{2}}\eta(N)_{\frac{1}{2}}]_{1,1}\eta(\bar{K})_{\frac{1}{2},-\frac{1}{2}}
 \nonumber\\
 &+\sqrt{\frac{1}{3}}[\eta(N)_{\frac{1}{2}}\eta(N)_{\frac{1}{2}}]_{1,0}\eta(\bar{K})_{\frac{1}{2},\frac{1}{2}}.\label{eq:ratio}
\end{align}
where the first (second) term corresponds to $K^{-}pp$ ($\bar{K}^0pn$).
This leads to the ratio of the probabilities 
of the $K^-pp$ and $\bar{K}^0pn$ components as
$P_{K^-}:P_{\bar{K}^0}= 2:1$.
The obtained $P_{\bar{K}N}^{I}$s in Tables~\ref{tab:Kpp} and \ref{tab:Kpn}
well satisfy these relations.
The small deviations 
from the ideal ratios
$2:1$ and $3:1$ come from 
the contributions of isospin-singlet $NN$ component with
odd wave and the Coulomb
interaction that induces the isospin mixing.

Because the Coulomb interaction is included in our formalism,
the energy splitting of the two members of the isospin doublet, $K^-pp$-$\bar{K}^0pn$ and $K^-pn$-$\bar{K}^0nn$, appears.
The splitting between these two systems is very
small, $\Delta B=B(K^-pp$-$\bar{K}^0pn)-B(K^-pn$-$\bar{K}^0nn) \sim 0.5$
MeV. There are two attractive
and one repulsive Coulombic pairs
in the $K^-pp$ channel, and one attractive
Coulombic pairs in the 
$K^-pn$ channel. Because $P_{K^{-}}=0.65$ in $K^{-}pp$-$\bar{K}^{0}pn$ is much larger than $P_{K^{-}}=0.38$ in $K^{-}pn$-$\bar{K}^{0}nn$, the Coulomb interaction
affects more attractively the $K^-pp$-$\bar{K}^0pn$ system 
than the $K^-pn$-$\bar{K}^0nn$ system.

Next, we show the particle density distributions of 
the
$K^-pp$-$\bar{K}^0pn$ system.
Figure~\ref{fig:K-pp_N_dist} plots the density distributions
of nucleons $\rho_N^{N\text{cm}}(r)$ defined in Eq.~\eqref{eq:rhoNcm}.
The density distribution of the deuteron is also plotted for comparison.
The central nucleon density $(r\lesssim 0.3$ fm) is suppressed due to the repulsive core of the nuclear force employed.
The nuclear system in the $\bar{K}NN$ system becomes more
compact than in the deuteron.
The shrinkage effect of nucleons with 
Type II  is slightly stronger than that with Type I. 
This is because the $\bar{K}N$
interactions with
Type II is more attractive than those with Type I
due to different $\delta\sqrt{s}$.
In Fig.~\ref{fig:K-pp_dist}, we also show the nucleon and antikaon
density distribution $\rho_N$ and
 $\rho_{\bar{K}}$ 
 defined in Eqs.~\eqref{eq:rhoN} and \eqref{eq:rhoK}.
The nucleon density is suppressed at around the origin,
while the antikaon density distribution is not suppressed since there are no
repulsive core in the $\bar{K}N$ potential.

We also perform the same calculations by employing the phenomenological
potential model, so-called Akaishi-Yamazaki
(AY) potential~\cite{Akaishi:2002bg,Yamazaki:2007cs}.
The results are shown in Tables~\ref{tab:Kpp} and \ref{tab:Kpn} and Figs.~\ref{fig:K-pp_N_dist} and \ref{fig:K-pp_dist}.
The AY potential is more attractive than the SIDDHARTA potential in the
subthreshold energy region.
The binding energies of the 
$\bar{K}NN$ systems are approximately twice of
 those with the SIDDHARTA potential, and the decay widths are
$\Gamma\sim 62$~MeV.
The particle density distributions become more compact, and the
rms radii are about $0.85$ times smaller than those for
the SIDDHARTA potential. Our results for AY potential are
comparable with the results in Ref.~\cite{Yamazaki:2007cs}.

\subsection{Structure of $\bar{K}NNN$ quasi-bound state}

\begin{table*}[tb]
\begin{center}
\caption{Properties of the calculation for $^3_{\bar{K}}\text{H}$ system
 with $J^\pi=1/2^-$.}
{\tabcolsep = 2.7mm
  \begin{tabular}{cccccccccc} \hline\hline
   \multicolumn{4}{c}{$^3_{\bar{K}}\text{H}~(J^\pi=1/2^-)$} \\\hline
   Model& \multicolumn{2}{c}{SIDDHARTA}&AY\\
   & Type I  & Type II &  \\\hline
  $B$ [MeV]     &$45.3$ & $49.7$  &$72.6$    \\
  $\Gamma$ [MeV] &$25.5$ & $69.4$ &$78.6$       \\
  $\delta \sqrt{s}$ [MeV]     &$-70.4-i20.7$ & $-26.1-i18.6$   &  \\
  $P_{K^-}$ &$0.53$ & $0.53$  &$0.51$      \\
  $P_{\bar{K}^0}$ &$0.47$ & $0.47$  &$0.49$      \\
  $\sqrt{\langle r_{NN}^2\rangle}$ [fm] &$1.99$ & $1.90$  &$1.87$  \\
  $\sqrt{\langle r_{\bar{K}N}^2\rangle}$ [fm] &$1.79$ & $1.68$  &$1.63$  \\
  $\sqrt{\langle r_N^{2}\rangle}$ [fm] &$1.17$ & $1.11$  &$1.09$     \\
  $\sqrt{\langle r_{\bar{K}}^{2}\rangle}$ [fm] &$1.17$ & $1.08$  &$1.03$    \\
  $\langle T\rangle^{K^-}_G$ [MeV] &$114+i17.4$ & $126+i42.2$  &$107+i27.6$    \\
  $\langle V\rangle^{K^-}_G$ [MeV] &$-118-i22.4$ & $-135-i54.0$  &$-114-i44.6$    \\
  $\langle T\rangle_G^{\bar{K}^0}$ [MeV] &$103+i15.9$ & $113+i39.8$  &$101+i26.1$    \\
  $\langle V\rangle_G^{\bar{K}^0}$ [MeV] &$-105 -i20.2$ & $-118-i50.0$  &$-107-i42.1$    \\
  $2\langle V\rangle_G^{K^-\bar{K}^0}$ [MeV] &$-39.0-i3.56$ & $-36.0-i12.7$  &$-59.6-i6.33$    \\
  $P_{\bar{K}N}^{I=0}$  &$0.50$ & $0.50$  &$0.50$    \\
  $P_{\bar{K}N}^{I=1}$  &$0.50$ & $0.50$  &$0.50$    \\
 \hline\hline
\end{tabular}  }
\label{tab:Kppn} 
\end{center}
\end{table*}
\begin{figure*}[tbh]
 \includegraphics[width=0.48\textwidth,clip]{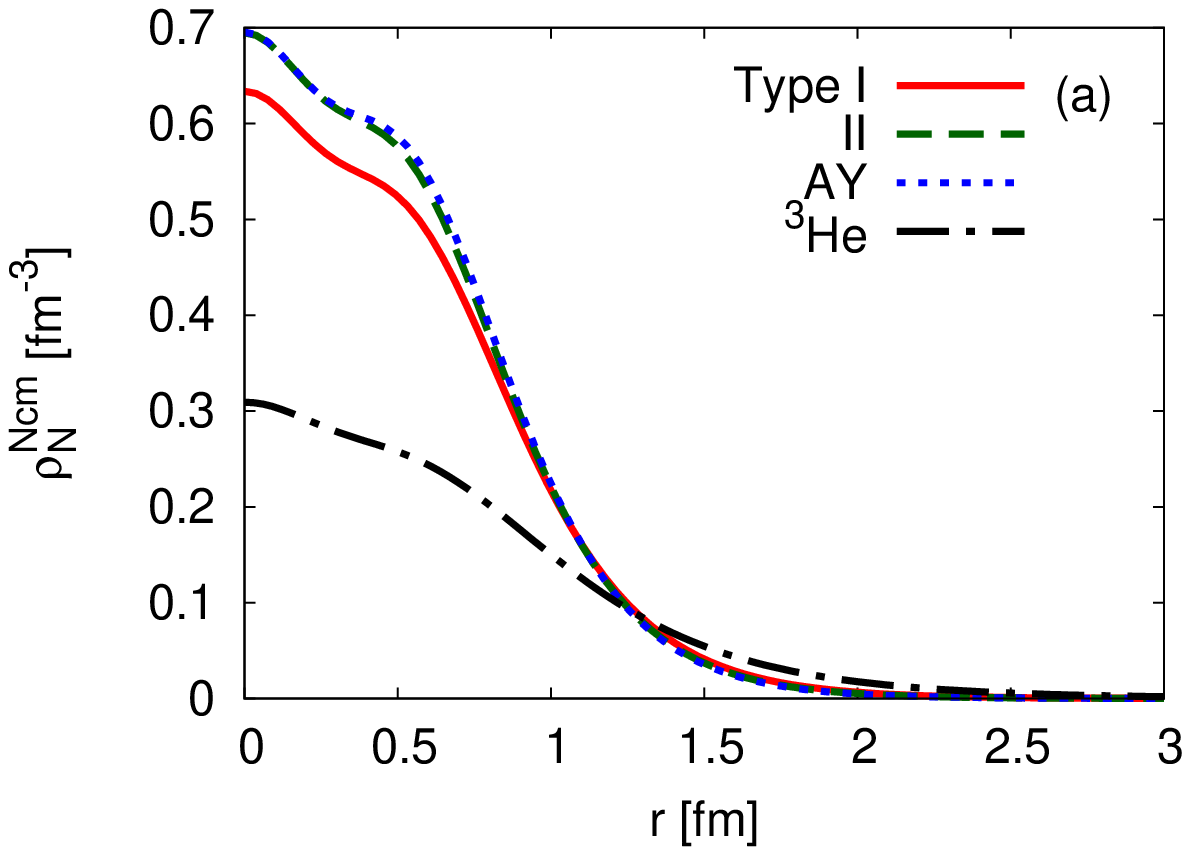}
 \includegraphics[width=0.48\textwidth,clip]{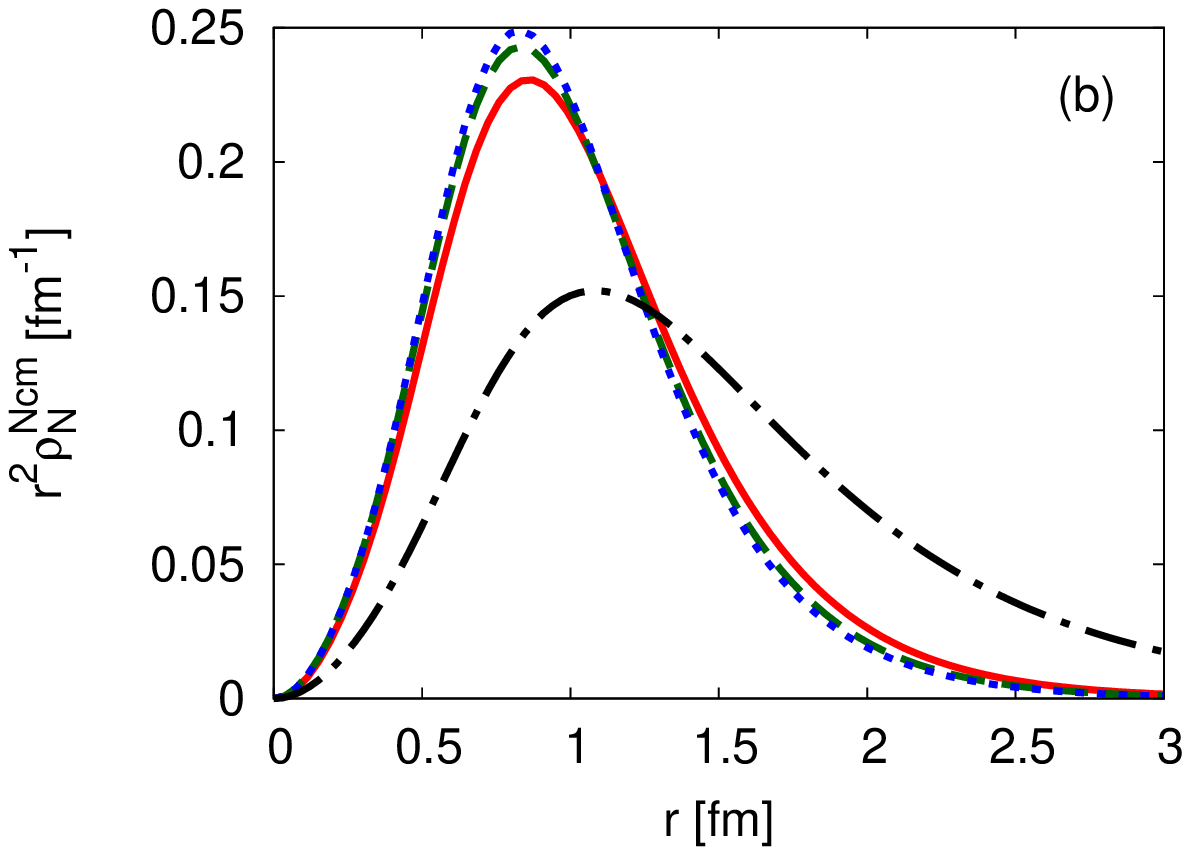}
 \caption{(Color online) Same as Fig.~\ref{fig:K-pp_N_dist}
   but for  $^3_{\bar{K}}\text{H}$ system.
 The nucleon density
 distributions for $^3$He is plotted for comparison.}
 \label{fig:K-ppn_N_dist}
\end{figure*}
\begin{figure*}[tbh]
 \includegraphics[width=0.48\textwidth,clip]{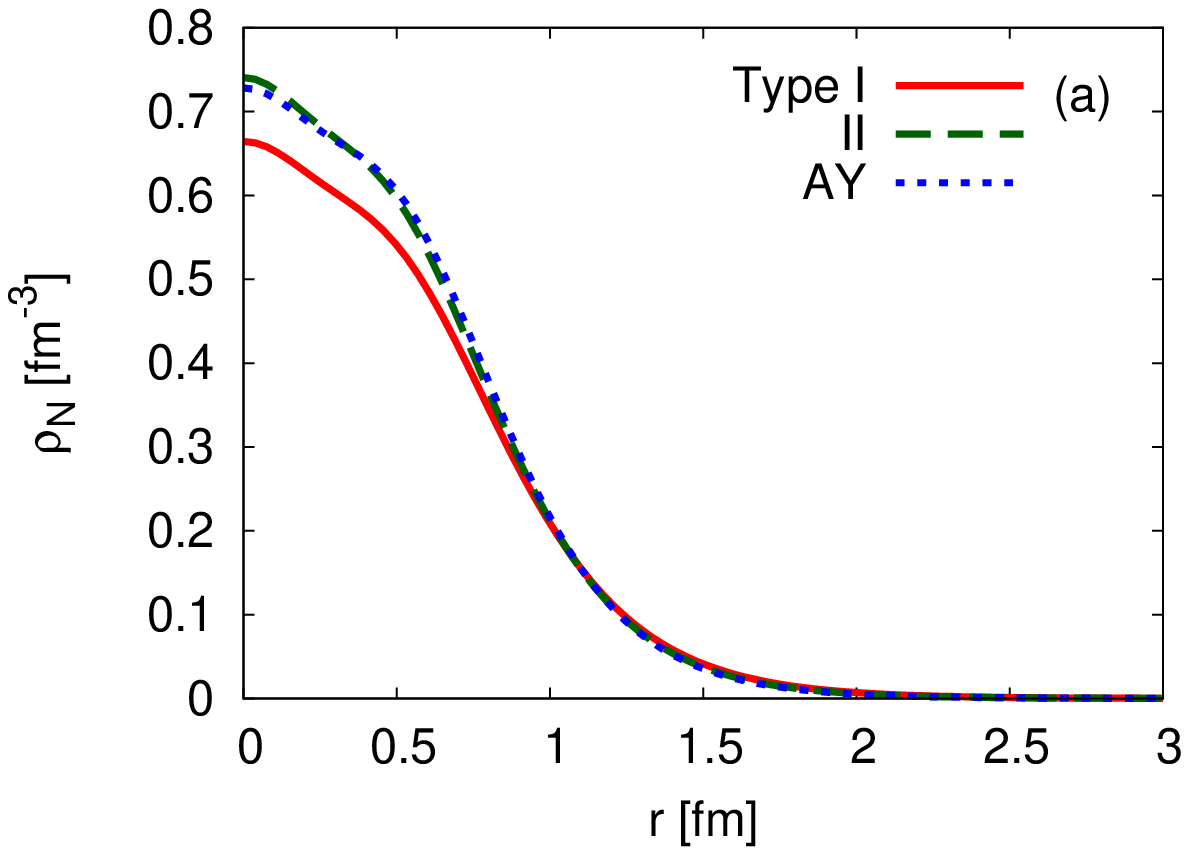}
 \includegraphics[width=0.48\textwidth,clip]{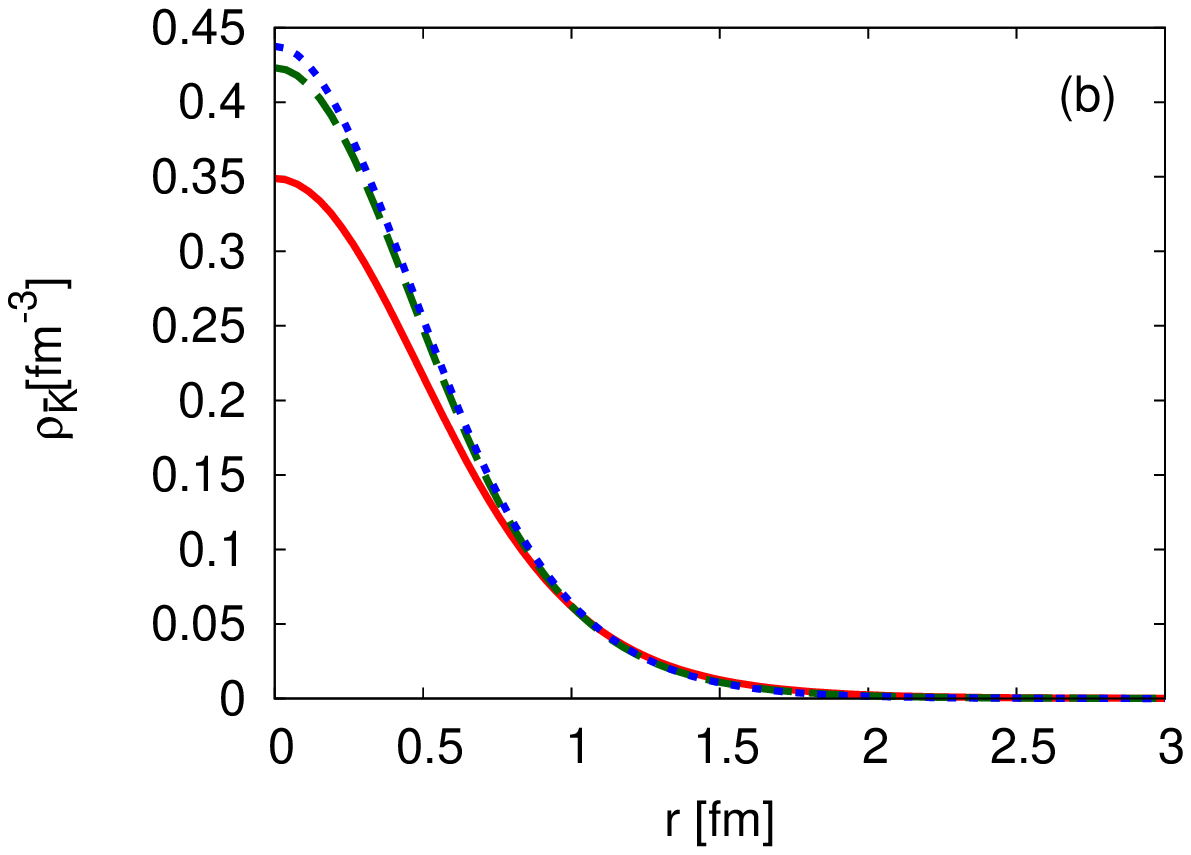}
 \caption{(Color online) Same as Fig.~\ref{fig:K-pp_dist} but
   for  $^3_{\bar{K}}\text{H}$ system.
 }
 \label{fig:K-ppn_dist}
\end{figure*}

Next, we investigate the structure of the four-body system,
strange tribaryon $\bar{K}NNN$ system with $J^\pi=1/2^-$.
We find a quasi-bound state in the $K^-ppn$-$\bar{K}^0pnn$ 
coupled system ($\equiv~^3_{\bar{K}}\text{H}$).
Our results are listed in Table~\ref{tab:Kppn}.
We also investigate the $K^-ppp$-$\bar{K}^0ppn$ coupled system
($\equiv~^3_{\bar{K}}\text{He}$) with $J^\pi=1/2^-$,
but we do not find any states below the $(\bar{K}NN)+N$ threshold.

We see a similar trend, shown in the three-body sector,
in the dependence of the choice of the two-body energy (Types I and II).
For Type II the real part of $\delta\sqrt{s}$ is less than a half
of those with Type I. 
Therefore the $\bar{K}N$ attraction with 
Type II is stronger than that with Type~I, 
and the binding energy with Type II is larger than that with Type I. 
The decay width with Type II ($\sim 69$ MeV) becomes
around three times as large as that with Type I 
($\sim 26$ MeV).
The obtained binding energies by using the SIDDHARTA potential are
$15$-$ 20$ MeV larger than the values obtained in
Ref.~\cite{Barnea:2012qa}.
As discussed in the three-body sector, this difference mainly comes from
the different treatment of the imaginary part of the potential as well
as the difference of the potential model.
Since the number of $\bar{K}N$ pairs is larger,
the effects of those differences are stronger in the
four-body systems than those in the three-body systems.

The rms distances
$\sqrt{\langle r_{NN}^2 \rangle}$, $\sqrt{\langle r_{\bar{K}N}^2 \rangle}$,
$\sqrt{\langle r_{N}^2 \rangle}$, and
$\sqrt{\langle r_{K}^2 \rangle}$ with 
Type II are slightly smaller than
those with Type I, in accordance with the larger binding.
The probabilities of finding the
$K^-ppn$ ($P_{K^-}$) and $\bar{K}^0pnn$
($P_{\bar{K}^0}$) channels are not sensitive to the choice of the two-body energies.
A large contribution to the real part from $2\langle V\rangle_G^{K^-\bar{K}^0}$ indicates that the
$K^-$-$\bar{K}^0$ channel coupling is essential for gaining the
binding energy for four-body systems,
while the diagonal channels also give contributions to
the decay width.

The $K^-ppn$ and $\bar{K}^0pnn$ channels are isospin mirror states.
Therefore, the probabilities of 
these two components follow 
$P_{K^-}:P_{\bar{K}^0}= 1:1$ with 
the isospin symmetric $\bar{K}N$ and
$NN$ interactions. 
In fact, the numerical results in Table~\ref{tab:Kppn} are consistent with this expectation within a small isospin mixing by the Coulomb interaction.
There are two attractive
and one repulsive Coulombic pairs in the $K^-ppn$ channel, while there are no
Coulomb interacting pair in the $\bar{K}^0pnn$ channel.
The Coulomb interaction in total affects attractive in 
the $K^-ppn$ channel,
and the $P_{K^-}$ becomes slightly larger
than the $P_{\bar{K}^0}$.

Figure~\ref{fig:K-ppn_N_dist} displays
$\rho_N^{N\text{cm}}(r)$ and
$r^2\rho_N^{N\text{cm}}(r)$ of the $^3_{\bar{K}}\text{H}$ system.
In the $\bar{K}NNN$ system, the nuclear system  becomes more
compact,
and the central density becomes about two times larger than that in the $^3$He.
Since the $\bar{K}N$
interaction with 
Type II is more attractive than 
that with Type I 
due to small magnitude of the real part of $\delta\sqrt{s}$,
the shrinkage effect of nucleons with 
Type II is slightly stronger than
that with Type I. 
In Fig.~\ref{fig:K-ppn_dist}, we show the nucleon and antikaon
density distribution, $\rho_N$ and
 $\rho_{\bar{K}}$.
Those density distributions are similar to each other,
and the antikaon rms radius
$\sqrt{\langle r_{\bar{K}}^{2}\rangle}$ is well comparable with the nucleon
rms radius $\sqrt{\langle r_{N}^{2}\rangle}$.
The antikaon moves in the whole region of the nuclear system
in order to gain the energy from the strong $\bar{K}N$ interaction.

With the AY potential,
the binding energy of the 
$\bar{K}NNN$ system is about 30 MeV larger
than the binding energies with the SIDDHARTA potential, and the decay width is
$\Gamma\sim 79$~MeV.
The particle density distributions and the
rms radii are similar to those obtained with the SIDDHARTA potential with 
Type II.
It is now clear that the deeply bound ($B>100$ MeV in
Ref.~\cite{Yamazaki:2002uh,Dote:2002db}) and high density (8.2 times of the normal density shown in Ref.~\cite{Dote:2002db}) $\bar{K}NNN$ state is not realized in the accurate few-body calculation. Such an extreme result can be regarded as an artifact due to the approximated treatment of the few-body systems.
We, however, emphasize that the existence of the bound state is confirmed also with the realistic SIDDHARTA $\bar{K}N$ interaction, and the central density of nucleons can be about two times larger than that in the $^3$He.

\subsection{Structure of $\bar{K}NNNN$ quasi-bound state}

\begin{table*}[tb]
\begin{center}
\caption{Properties of the calculation for $^4_{\bar{K}}\text{He}$ system
 with $J^\pi=0^-$.}
{\tabcolsep = 2.7mm
  \begin{tabular}{cccccccccc} \hline\hline
   \multicolumn{4}{c}{$^4_{\bar{K}}\text{He}~(J^\pi=0^-)$} \\\hline
   Model& \multicolumn{2}{c}{SIDDHARTA}&AY\\
   & Type I  & Type II &  \\\hline
  $B$ [MeV]     &$67.9$ & $72.7$  &$85.2$    \\
  $\Gamma$ [MeV] &$28.3$ & $74.1$ &$86.5$       \\
  $\delta \sqrt{s}$ [MeV]     &$-67.6-i23.0$ & $-18.4-i15.0$   &  \\
  $P_{K^-}$ &$0.08$ & $0.06$  &$0.16$      \\
  $P_{\bar{K}^0}$ &$0.92$ & $0.94$  &$0.84$      \\
  $\sqrt{\langle r_{NN}^2\rangle}$ [fm] &$1.98$ & $1.91$  &$2.07$  \\
  $\sqrt{\langle r_{\bar{K}N}^2\rangle}$ [fm] &$1.83$ & $1.72$  &$1.81$  \\
  $\sqrt{\langle r_N^{2}\rangle}$ [fm] &$1.22$ & $1.18$  &$1.27$     \\
  $\sqrt{\langle r_{\bar{K}}^{2}\rangle}$ [fm] &$1.22$ & $1.12$  &$1.14$    \\
  $\langle T\rangle^{K^-}_G$ [MeV] &$32.6+i6.75$ & $26.7+i16.2$  &$50.3+i7.22$    \\
  $\langle V\rangle^{K^-}_G$ [MeV] &$-25.1-i6.74$ & $-20.0-i15.9$  &$-42.3-i12.0$    \\
  $\langle T\rangle_G^{\bar{K}^0}$ [MeV] &$214+i27.8$ & $240+i66.0$  &$183+i52.7$    \\
  $\langle V\rangle_G^{\bar{K}^0}$ [MeV] &$-265 -i38.4$ & $-300-i94.3$  &$-232-i88.3$    \\
  $2\langle V\rangle_G^{K^-\bar{K}^0}$ [MeV] &$-24.8-i3.66$ & $-20.2-i9.13$  &$-43.9-i2.82$    \\
  $P_{\bar{K}N}^{I=0}$  &$0.28$ & $0.27$  &$0.31$    \\
  $P_{\bar{K}N}^{I=1}$  &$0.72$ & $0.73$  &$0.69$    \\
 \hline\hline
\end{tabular}  }
\label{tab:Kpppn} 
\end{center}
\end{table*}
\begin{table*}[tb]
\begin{center}
\caption{Properties of the calculation for $^4_{\bar{K}}\text{H}$ system
 with $J^\pi=0^-$.}
{\tabcolsep = 2.7mm
  \begin{tabular}{cccccccccc} \hline\hline
   \multicolumn{4}{c}{$^4_{\bar{K}}\text{H}~(J^\pi=0^-)$} \\\hline
   Model& \multicolumn{2}{c}{SIDDHARTA}&AY\\
   & Type I  & Type II &  \\\hline
  $B$ [MeV]     &$69.6$ & $75.5$  &$87.4$    \\
  $\Gamma$ [MeV] &$28.0$ & $74.5$ &$87.2$       \\
  $\delta \sqrt{s}$ [MeV]     &$-68.7-i22.4$ & $-19.1-i14.9$   &  \\
  $P_{K^-}$ &$0.93$ & $0.94$  &$0.86$      \\
  $P_{\bar{K}^0}$ &$0.07$ & $0.06$  &$0.14$      \\
  $\sqrt{\langle r_{NN}^2\rangle}$ [fm] &$1.96$ & $1.89$  &$2.04$  \\
  $\sqrt{\langle r_{\bar{K}N}^2\rangle}$ [fm] &$1.82$ & $1.71$  &$1.79$  \\
  $\sqrt{\langle r_N^{2}\rangle}$ [fm] &$1.21$ & $1.17$  &$1.26$     \\
  $\sqrt{\langle r_{\bar{K}}^{2}\rangle}$ [fm] &$1.21$ & $1.11$  &$1.13$    \\
  $\langle T\rangle^{K^-}_G$ [MeV] &$216+i28.1$ & $244+i66.6$  &$188+i53.4$    \\
  $\langle V\rangle^{K^-}_G$ [MeV] &$-269-i39.1$ & $-306-i95.7$  &$-241-i90.0$    \\
  $\langle T\rangle_G^{\bar{K}^0}$ [MeV] &$30.5+i5.50$ & $25.1+i14.7$  &$47.0+i6.54$    \\
  $\langle V\rangle_G^{\bar{K}^0}$ [MeV] &$-23.0 -i5.54$ & $-18.5-i14.2$  &$-38.7-i10.9$    \\
  $2\langle V\rangle_G^{K^-\bar{K}^0}$ [MeV] &$-24.4-i3.00$ & $-19.9-i8.71$  &$-42.7+i2.64$    \\
  $P_{\bar{K}N}^{I=0}$  &$0.28$ & $0.27$  &$0.30$    \\
  $P_{\bar{K}N}^{I=1}$  &$0.72$ & $0.73$  &$0.70$    \\
 \hline\hline
\end{tabular}  }
\label{tab:Kppnn} 
\end{center}
\end{table*}
\begin{figure*}[tbh]
 \includegraphics[width=0.48\textwidth,clip]{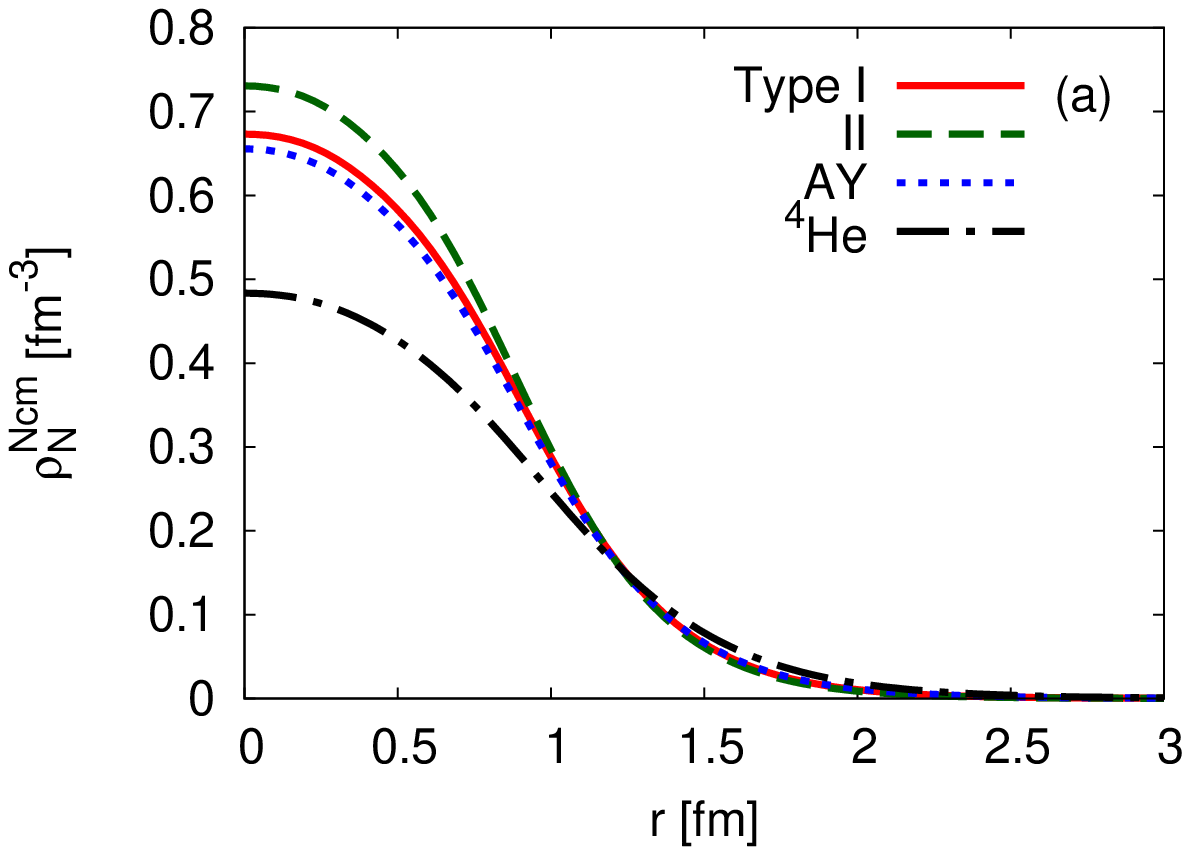}
 \includegraphics[width=0.48\textwidth,clip]{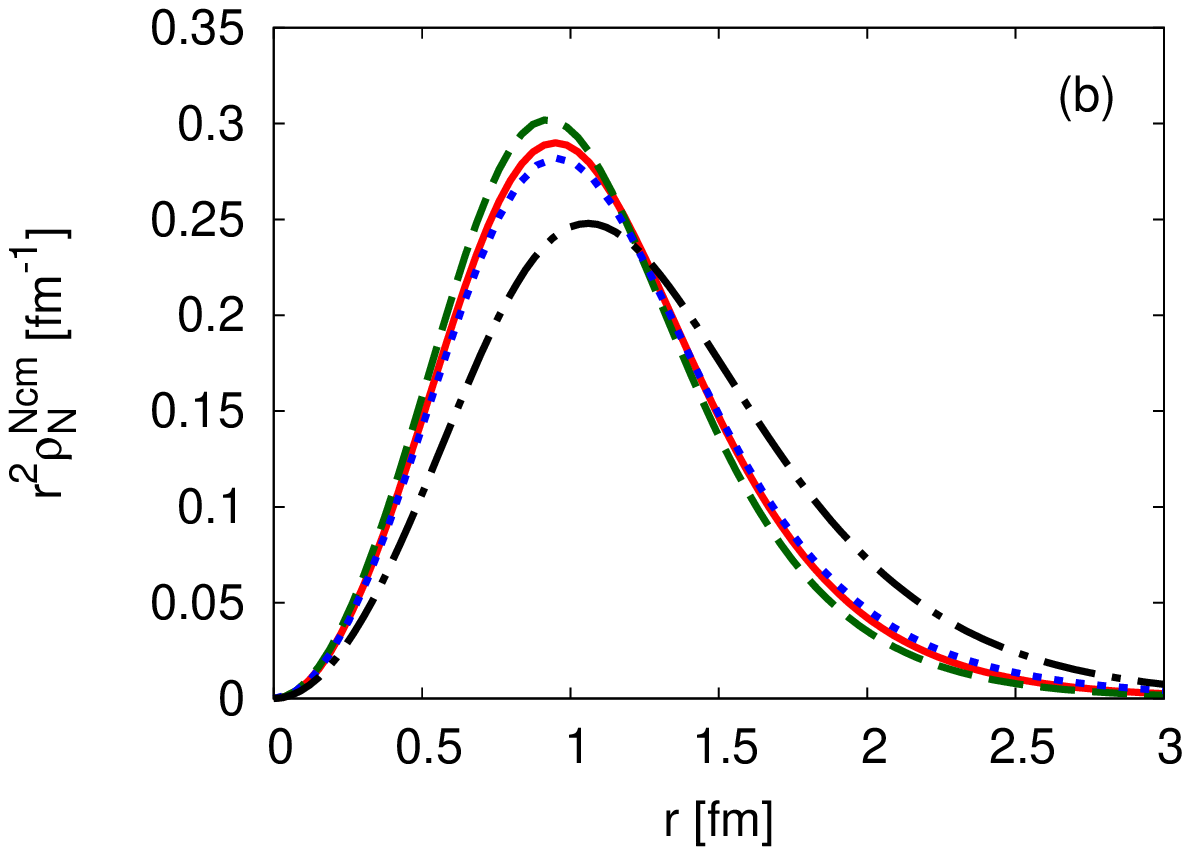}
 \caption{(Color online) Same as Fig.~\ref{fig:K-pp_N_dist}
   but for  $^4_{\bar{K}}\text{H}$ system.
   The nucleon density distribution for $^4$He is also plotted for
comparison.}
 \label{fig:K-ppnn_N_dist}
\end{figure*}
\begin{figure*}[tbh]
 \includegraphics[width=0.48\textwidth,clip]{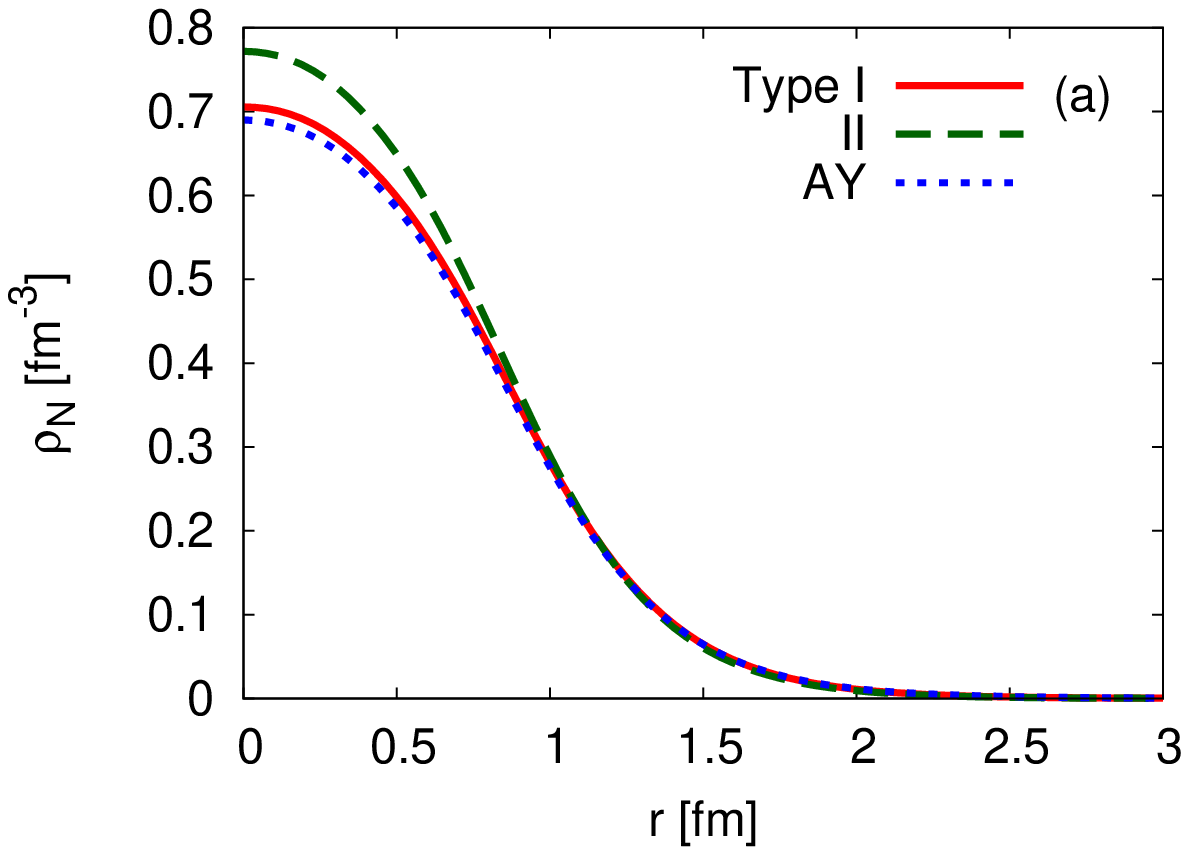}
 \includegraphics[width=0.48\textwidth,clip]{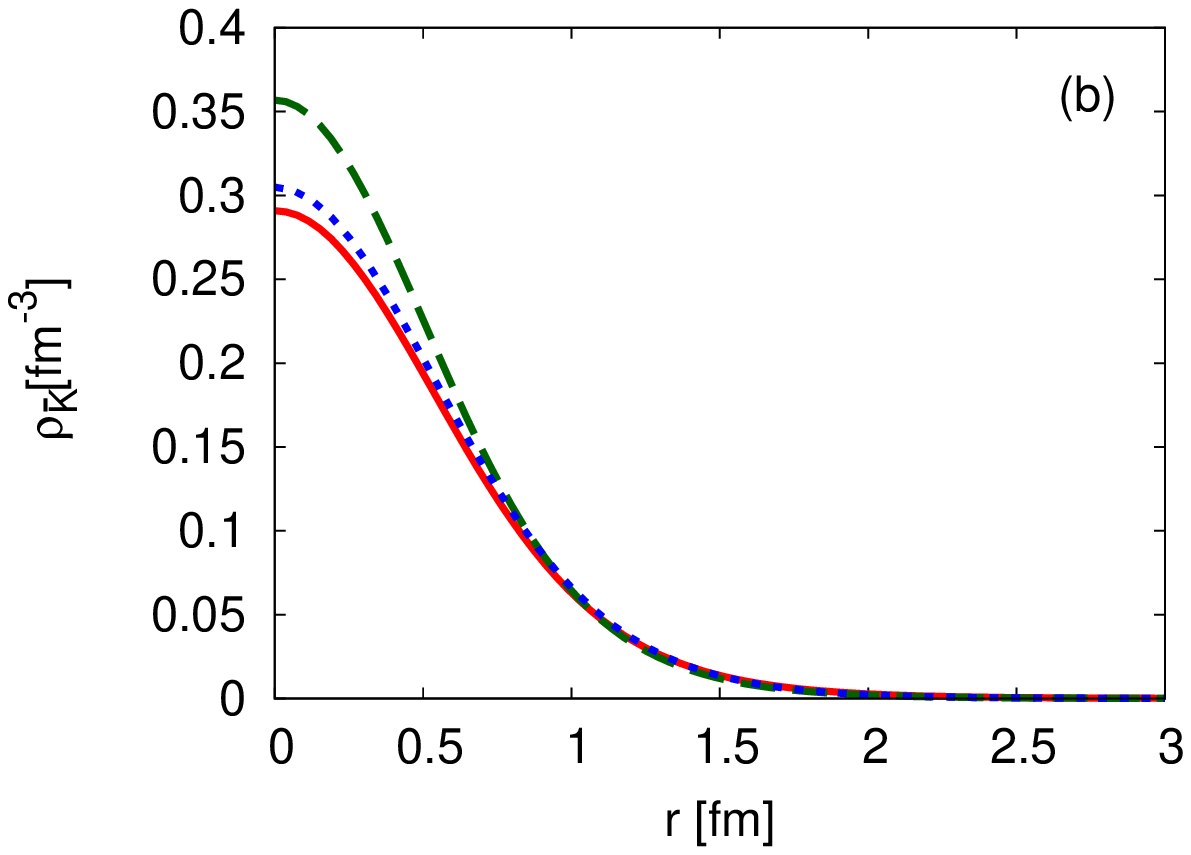}
 \caption{(Color online) Same as Fig.~\ref{fig:K-pp_dist}
   but for  $^4_{\bar{K}}\text{H}$ system.
}
 \label{fig:K-ppnn_dist}
\end{figure*}

Next, we investigate the structure of the five-body systems,
strange tetrabaryon $\bar{K}NNNN$ systems with $J^\pi=0^-$.
We find 
isospin-doublet quasi-bound states in the
$K^-pppn$-$\bar{K}^0ppnn$ ($\equiv~^4_{\bar{K}}\text{He}$) and
$K^-ppnn$-$\bar{K}^0pnnn$ ($\equiv~^4_{\bar{K}}\text{H}$) systems.
Our results are listed in Tables~\ref{tab:Kpppn}
and \ref{tab:Kppnn}.

The binding energy with
Type II is 5 MeV larger than that with Type I.
The decay width with Type II ($\sim 75$ MeV)
 is about three times larger than that with  
Type I ($\sim 28$ MeV). The rms distances
$\sqrt{\langle r_{NN}^2 \rangle}$, $\sqrt{\langle r_{\bar{K}N}^2 \rangle}$,
$\sqrt{\langle r_{N}^2 \rangle}$, and
$\sqrt{\langle r_{K}^2 \rangle}$ with 
Type II are smaller than those with Type I.
The probabilities $P_{K^-}$ and
$P_{\bar{K}^0}$ are not sensitive to the choice of Types I and II. 
In contrast to these features, the energy decomposition of the $\bar{K}NNNN$ exhibits different characteristics from the three- and four-body systems. The incomplete cancellation of the kinetic energy and potential energy in the dominant component ($\langle T\rangle^{\bar{K}^0}_G+\langle K\rangle^{\bar{K}^0}_G$ in $^4_{\bar{K}}\text{He}$
and $\langle T\rangle^{K^-}_G+\langle K\rangle^{K^-}_G$ in $^4_{\bar{K}}\text{H}$) leaves sizable contributions to the binding energy, which are comparable with the off-diagonal components $2\langle V\rangle_G^{K^-\bar{K}^0}$.
For the decay widths, the diagonal channels are also important  as in the 
four-body systems.

From the results of $P_{K^{-}}$ and $P_{\bar{K}^{0}}$, we see that the 
dominant component in the $^4_{\bar{K}}\text{He}$
($^4_{\bar{K}}\text{H}$) system is the $\bar{K}^0ppnn$ ($K^-ppnn$)
channel,
although the $K^-pppn$ ($\bar{K}^0pnnn$) channel contains 
more $\bar{K}N$ $I=0$
components than the $\bar{K}^0ppnn$ ($K^-ppnn$) channel.
This is because the nucleon contribution of the $\bar{K}^0ppnn$ ($K^-ppnn$)
channel, which can form an 
$\alpha$-particle configuration giving the binding energy
about 30 MeV, is larger than the nucleon contribution of
the $K^-pppn$ ($\bar{K}^0pnnn$) channel, and thus the $\bar{K}^0ppnn$ ($K^-ppnn$)
channel is favored.
It is also for this reason that the $\bar{K}^0$ ($K^-$) diagonal
component gains the binding energy
in the $^4_{\bar{K}}\text{He}$ ($^4_{\bar{K}}\text{H}$) system, as we see above.

The Coulomb splitting between these two systems is larger than
that in the $\bar{K}NN$ systems, $\Delta
 B=B(^4_{\bar{K}}\text{He})-B(^4_{\bar{K}}\text{H}) \sim 2$ MeV.
 There are one repulsive Coulombic pair
 in the 
 $\bar{K}^0ppnn$ channel which
 is the dominant component of the $^4_{\bar{K}}\text{He}$ system, and two attractive
 and one repulsive Coulombic pairs
 in the 
 $K^-ppnn$ channel which is the dominant component of
the $^4_{\bar{K}}\text{H}$ system.
Therefore, Coulomb interaction is repulsive in 
the
 $^4_{\bar{K}}\text{He}$ system and attractive in
 the
$^4_{\bar{K}}\text{H}$ system,
and the Coulomb splitting becomes
 larger than that of the three-body systems.

Figures~\ref{fig:K-ppnn_N_dist} and \ref{fig:K-ppnn_dist} plot
the particle density distributions in the
$^4_{\bar{K}}\text{H}$ system.
The nucleons in the $\bar{K}NNNN$ system become more compact,
and the central density increases to 
about 1.3-1.5 times higher than that in the $^4$He.
As in the
three- and four-body systems,
the shrinkage effect of nucleons with Type II is slightly stronger than
that with Type I. 
The antikaon density distribution is similar to the nucleon density distribution.

When we use the AY potential,
the binding energies of the $\bar{K}NNNN$ system are about 12-24 MeV larger
than those with the SIDDHARTA potential.
Because the quasi-bound state appears above the $\pi\Sigma NNN$ threshold, it has a sizable decay width of about
$\Gamma\sim 87$~MeV, in contrast to the narrow state predicted in Ref.~\cite{Akaishi:2002bg}.
The probability $P_{K^-}$ ($P_{\bar{K}^0}$) in the $^4_{\bar{K}}\text{He}$
($^4_{\bar{K}}\text{H}$) system 
becomes larger than the result with the SIDDHARTA potential
because the AY potential model has more attractive $\bar{K}N$ $I=0$
interaction than that in the 
SIDDHARTA potential.
As a result,
$P_{\bar{K}N}^{I=0}$ is enhanced.
The rms radius of 
the antikaon is smaller than 
the corresponding result of the SIDDHARTA potential
with Type I, 
whereas the nucleon radius 
is slightly larger.

\subsection{Nuclear force dependence}
\begin{figure*}[tbh]
 \includegraphics[width=0.48\textwidth,clip]{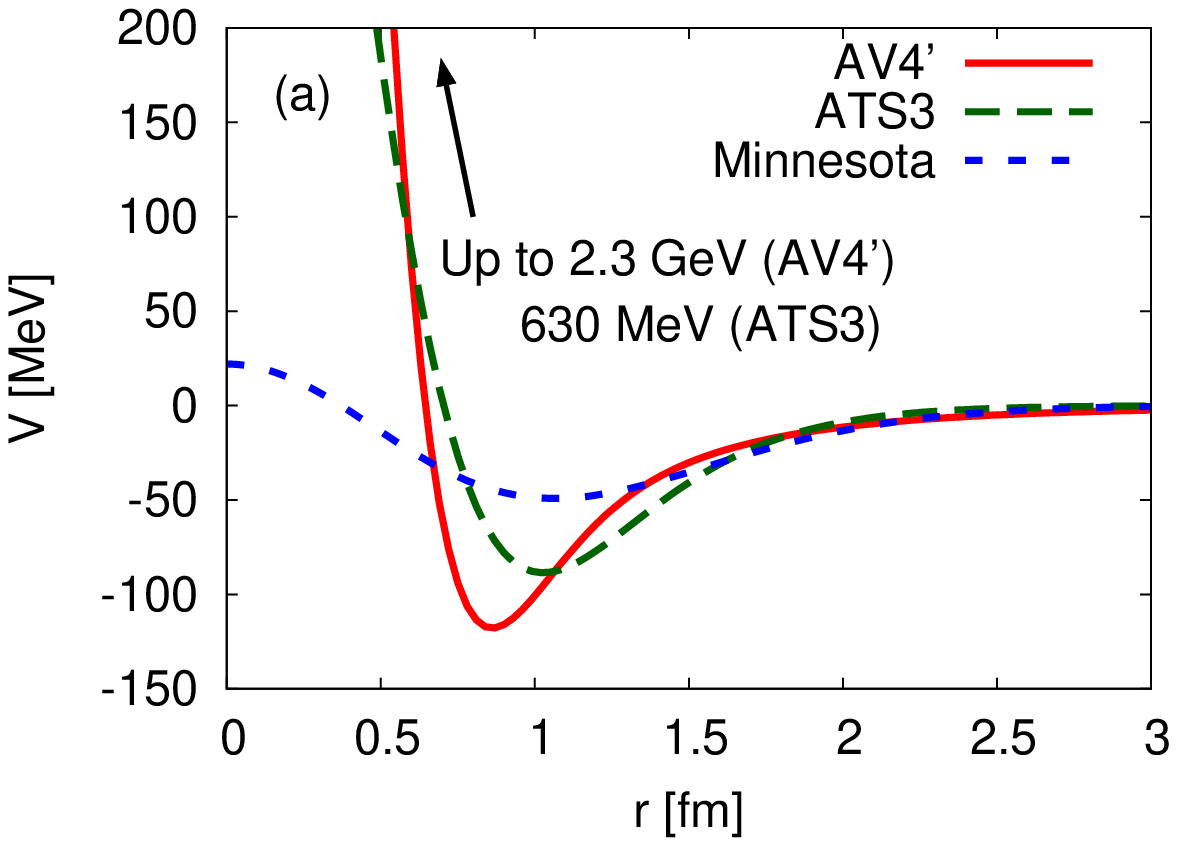}
 \includegraphics[width=0.48\textwidth,clip]{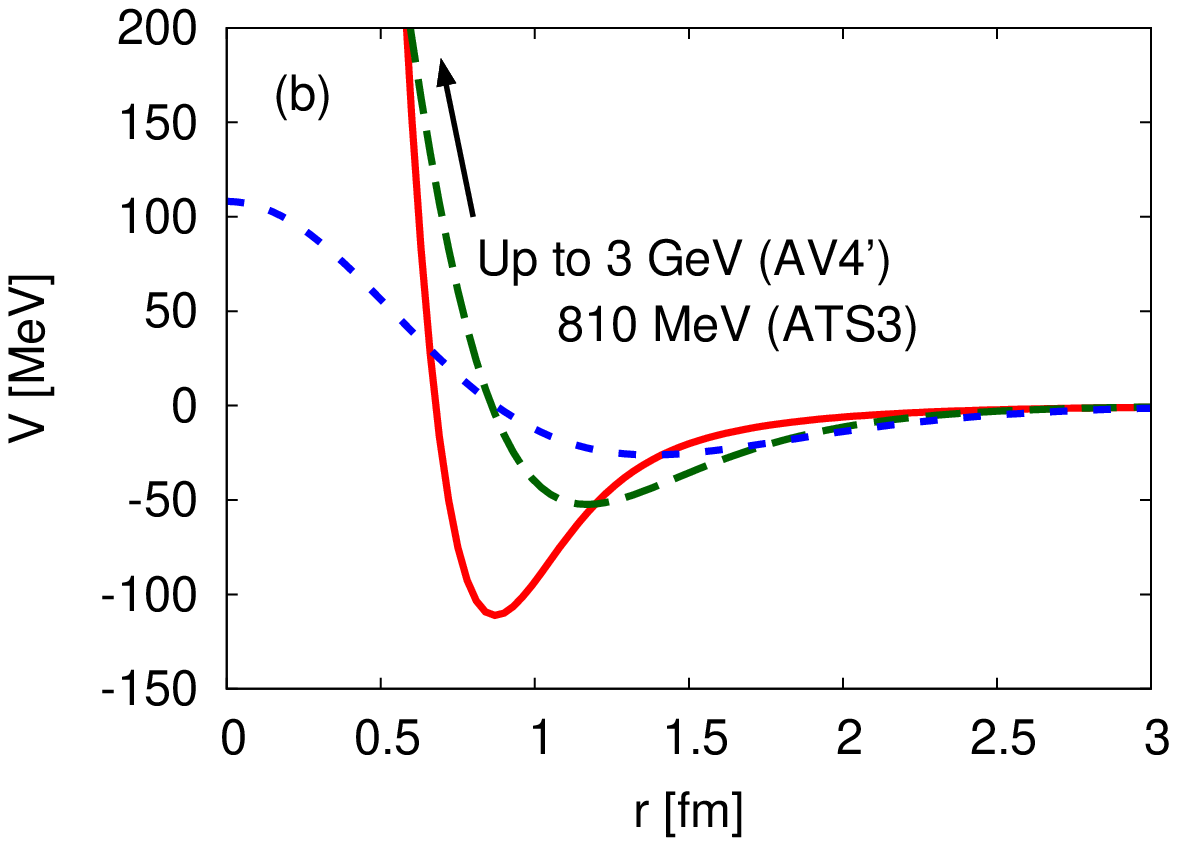}
 \caption{(Color online) (a) $^3S_1$ and (b)
 $^1S_0$ channels of the AV4', ATS3, and MN $NN$ interactions.
 }
 \label{fig:NN_int}
\end{figure*}
\begin{figure*}[tbh]
 \includegraphics[width=\textwidth,clip]{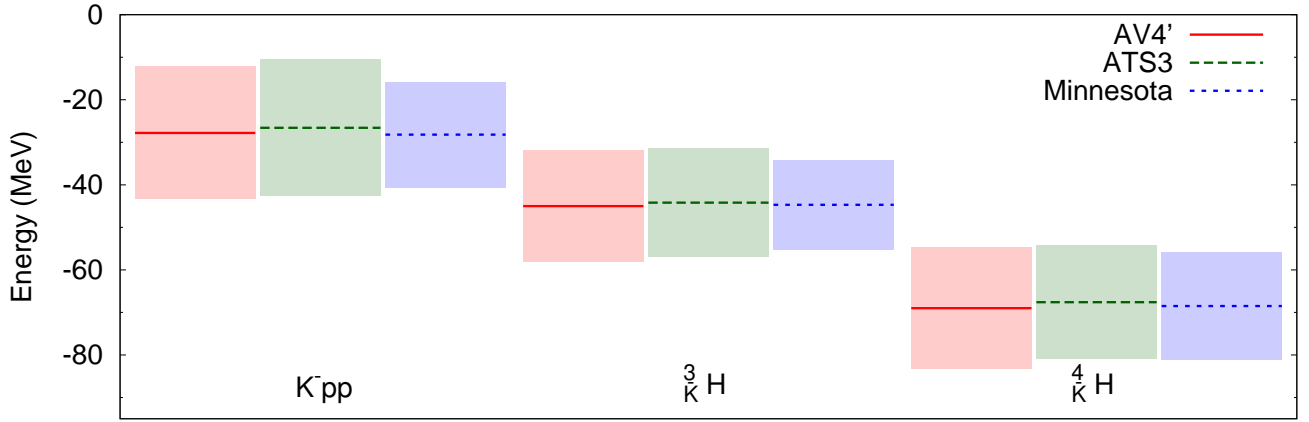}
 \caption{(Color online)
 Energy levels of $K^-pp~(J^\pi=0^-)$, $^3_{\bar{K}}\text{H}(J^\pi=1/2^-)$,
 and $^4_{\bar{K}}\text{H}(J^\pi=0^-)$ systems
 with AV4', ATS3, and MN potentials.
 The shaded widths represent the decay widths.
 SIDDHARTA potential Type I is employed for $\bar{K}N$ interaction.}
 \label{fig:energy_Ndep}
\end{figure*}
\begin{table*}[tb]
\begin{center}
 \caption{Total binding energies $B$ and decay widths $\Gamma$ of
 ordinary nuclei and kaonic nuclei.
 SIDDHARTA potential Type I is employed for $\bar{K}N$ interaction.
 }
{\tabcolsep = 2.7mm
  \begin{tabular}{cccccccc} \hline\hline
   & \multicolumn{3}{c}{$B$ [MeV]} &  & \multicolumn{3}{c}{$(B,~\Gamma)$ [MeV]} \\
   & AV4' & ATS3& MN &   & \multicolumn{1}{c}{AV4'}& \multicolumn{1}{c}{ATS3} & \multicolumn{1}{c}{MN}   \\
   \hline
  $^2$H     & $2.24$ &  $2.22$ & $2.20$ & $K^-pp$-$\bar{K}^0pn$ & $(27.9,~30.9)$ & $(26.6,~32.0)$& $(28.4,~23.9)$   \\
  $^3$He    & $8.33$ &  $8.11$ & $7.72$ & $^3_{\bar{K}}$H & $(45.3,~25.5)$ & $(44.6,~25.0)$& $(45.7,~19.3)$   \\
  $^4$He    & $32.1$ &  $30.8$ & $30.0$ & $^4_{\bar{K}}$H & $(69.6,~28.0)$ & $(68.7,~26.0)$& $(68.6,~24.6)$   \\
 \hline\hline
\end{tabular}  }
\label{tab:n-dep} 
\end{center}
\end{table*}

\begin{figure*}[tbh]
 \includegraphics[width=\textwidth,clip]{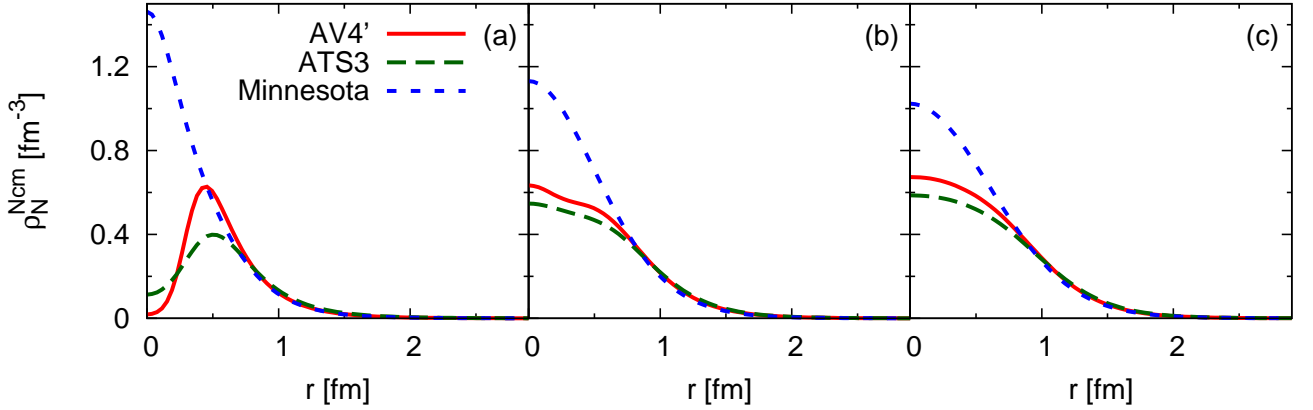}
 \caption{(Color online)
   Nucleon density distribution measured from the center-of-mass
     system of nucleons 
     for (a) $K^-pp$, (b) $^3_{\bar{K}}\text{H}$, and (c) $^4_{\bar{K}}\text{H}$ systems
     with the AV4', ATS3, and MN potentials.
 The SIDDHARTA potential with Type I
 is employed for $\bar{K}N$ interaction.}
 \label{fig:N_density_Ndep}
\end{figure*}
\begin{figure*}[tbh]
 \includegraphics[width=\textwidth,clip]{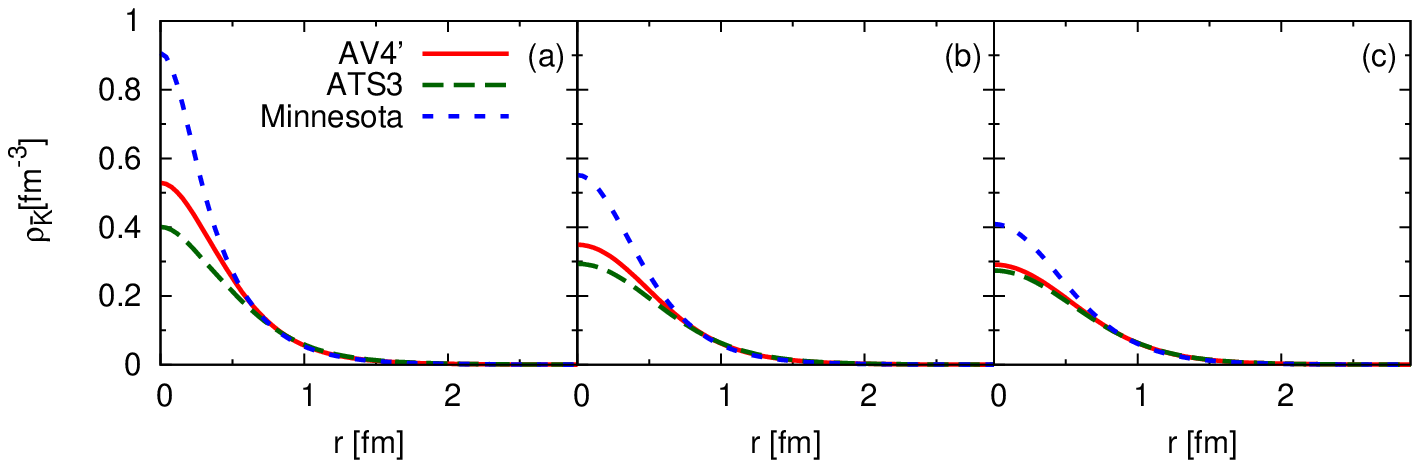}
 \caption{(Color online)
 Antikaon density distribution $\rho_{\bar{K}}$ in the center-of-mass system
 for (a) $K^-pp$, (b) $^3_{\bar{K}}\text{H}$, and (c) $^4_{\bar{K}}\text{H}$ systems.
 SIDDHARTA potential Type I is employed for $\bar{K}N$ interaction.}
 \label{fig:K_density_Ndep}
\end{figure*}

Here we discuss the $NN$ interaction dependence of our results by
comparing the results with the AV4'
potential and those with 
other $NN$ interaction models such as
Afnan-Tang S3 (ATS3) \cite{Afnan:1968zj} and Minnesota (MN) potential models~\cite{Thompson:1977zz}.
These potential models are often used
in studying light nuclei,
and well reproduce the binding energy of the $s$-shell nuclei.
It is noted that the strengths of the repulsive core are quite different
between these three models as displayed
in Fig.~\ref{fig:NN_int}.
The AV4' potential has the strongest
repulsive core, which is
comparable to the realistic nuclear forces such as the
  Argonne V18 potential model~\cite{Wiringa:1994wb}.
  The ATS3 potential model has also a strong short-range repulsion
  at around the origin.
The repulsive core of the MN potential is quite soft.
Since the $\bar{K}N$ interaction is strongly attractive, and
the kaonic nuclei become more compact than ordinary nuclei.
Therefore, there is a possibility that these different
repulsive cores affect the results of the kaonic nuclei.

Figure~\ref{fig:energy_Ndep}
displays
the binding energies and decay widths of the $K^-pp$-$\bar{K}^0pn$ 
($J^\pi=0^-$),
$_{\bar{K}}^3$H ($J^\pi=1/2^-$) and $_{\bar{K}}^4$H
($J^\pi=0^-$) systems with those three types of nuclear
forces.
Here, we use the SIDDHARTA potential as the $\bar{K}N$ potential, and the
energy dependence of the potential is determined by Type I.
The binding energies and decay widths are almost the same in those
with three nuclear forces as well as the binding energies
of ordinary $s$-shell nuclei without the antikaon
as listed in Table~\ref{tab:n-dep}.

The qualitative difference 
becomes apparent in the density distributions.
Figures~\ref{fig:N_density_Ndep} and \ref{fig:K_density_Ndep}
plot the nucleon and antikaon density distributions of the
$K^-pp$-$\bar{K}^0pn$ ($J^\pi=0^-$),
$^3_{\bar{K}}\text{H}$ ($J^\pi=1/2^-$)
and $^4_{\bar{K}}\text{H}$ ($J^\pi=0^-$) systems.
The density distributions with  
the AV4' potential model are similar to
those with the ATS3 model,
while the central densities with the 
MN potential with repulsive
core are significantly higher
than those for the other potential models.
Since the short-range repulsive core of the MN potential
  is not as strong as those of the other potential models,
  nucleons can come very close to each other due to
  the strong $\bar{K}N$ attraction.
In the $^3_{\bar{K}}\text{H}$ system,
the central density obtained with the MN potential model
becomes $\rho_N^{N\text{cm}}(r=0)\sim 1.2$ fm$^{-3}$,
  approximately two times larger than
those with the 
AV4' and ATS3 potential models.
The value is close to 
$1.4$ fm$^{-3}$ predicted
in Ref.~\cite{Dote:2002db,Dote:2003ac}
by using the antisymmetrized-molecular-dynamics
calculation with the effective treatment of the $\bar{K}N$ and
$NN$ interactions with the $g$-matrix.
Since the $\bar{K}N$ interaction is very strong,
the nucleons can be compressed too much with a soft core potential
and form such an unrealistically high density state.
Use of the realistic nuclear force is necessary
in order to avoid such an artificial solution.

\subsection{$\bar{K}NNNNN$ quasi-bound state}

For the six-body $\bar{K}NNNNN$ systems,
we could not find any
states below the strange tetrabaryon and a nucleon $(\bar{K}NNNN)+N$ threshold
energy in the $L=0$ state.
Since the $^5$He ground state is observed as a resonant state with $J^\pi=3/2^-$, the orbital angular momentum of the ground state is expected to be $L=1$.

  Investigation with $L>0$ states is possible by introducing
  the global 
  vectors that efficiently describe the rotational motion of
  the system with
  any $L^\pi$
  \cite{Varga:1995dm,suzuki1998stochastic,Suzuki:2008fg,Aoyama:2011ih},
  but this extension 
  is beyond the scope of this paper.

\subsection{Structure of $\bar{K}NNNNNN$ quasi-bound state}

\begin{table*}[tb]
\begin{center}
\caption{Properties of the calculation for $^6_{\bar{K}}\text{Li}$ system
 with $J^\pi=0^-$.}
{\tabcolsep = 2.7mm
  \begin{tabular}{cccccccccc} \hline\hline
   \multicolumn{4}{c}{$^6_{\bar{K}}\text{Li}~(J^\pi=0^-)$} \\\hline
   Model& \multicolumn{2}{c}{SIDDHARTA}&AY\\
   & Type I  & Type II &  \\\hline
  $B$ [MeV]     &$69.8$ & $79.7$  &$103$    \\
  $\Gamma$ [MeV] &$23.7$ & $75.6$ &$88.0$       \\
  $\delta \sqrt{s}$ [MeV]     &$-76.2-i18.1$ & $-15.1-i10.2$   &  \\
  $P_{K^-}$ &$0.70$ & $0.72$  &$0.64$      \\
  $P_{\bar{K}^0}$ &$0.30$ & $0.28$  &$0.36$      \\
  $\sqrt{\langle r_{NN}^2\rangle}$ [fm] &$2.80$ & $2.75$  &$2.57$  \\
  $\sqrt{\langle r_{\bar{K}N}^2\rangle}$ [fm] &$2.52$ & $2.40$  &$2.27$  \\
  $\sqrt{\langle r_N^{2}\rangle}$ [fm] &$1.82$ & $1.78$  &$1.66$     \\
  $\sqrt{\langle r_{\bar{K}}^{2}\rangle}$ [fm] &$1.61$ & $1.49$  &$1.43$    \\
  $\langle T\rangle^{K^-}_G$ [MeV] &$186+i19.3 $ & $213+i47.9 $  &$184+i34.4 $    \\
  $\langle V\rangle^{K^-}_G$ [MeV] &$-220 -i28.5$ & $-258 -i69.3$ &$-224 -i61.7$    \\
  $\langle T\rangle_G^{\bar{K}^0}$ [MeV] &$89.5+i5.08$ & $88.9+i26.5$  &$110+i17.5$    \\
  $\langle V\rangle_G^{\bar{K}^0}$ [MeV] &$-95.4-i8.43$ & $-97.6-i33.4$  &$-121-i31.7$    \\
  $2\langle V\rangle_G^{K^-\bar{K}^0}$ [MeV] &$-30.1+i0.703$ & $-25.6-i9.55$  &$-52.4-i2.59$    \\
  $P_{\bar{K}N}^{I=0}$  &$0.41$ & $0.41$  &$0.41$    \\
  $P_{\bar{K}N}^{I=1}$  &$0.59$ & $0.59$  &$0.59$    \\
 \hline\hline
\end{tabular}  }
\label{tab:Kppppnn} 
\end{center}
\end{table*}
\begin{table*}[tb]
\begin{center}
\caption{Properties of the calculation for $^6_{\bar{K}}\text{He}$ system
 with $J^\pi=0^-$.}
{\tabcolsep = 2.7mm
  \begin{tabular}{cccccccccc} \hline\hline
   \multicolumn{4}{c}{$^6_{\bar{K}}\text{He}~(J^\pi=0^-)$} \\\hline
   Model& \multicolumn{2}{c}{SIDDHARTA}&AY\\
   & Type I  & Type II &  \\\hline
  $B$ [MeV]     &$70.6$ & $80.0$  &$103$    \\
  $\Gamma$ [MeV] &$23.9$ & $75.5$ &$88.0$       \\
  $\delta \sqrt{s}$ [MeV]     &$-75.6-i18.3$ & $-14.9-i10.2$   &  \\
  $P_{K^-}$ &$0.41$ & $0.41$  &$0.40$      \\
  $P_{\bar{K}^0}$ &$0.59$ & $0.59$  &$0.60$      \\
  $\sqrt{\langle r_{NN}^2\rangle}$ [fm] &$2.79$ & $2.74$  &$2.56$  \\
  $\sqrt{\langle r_{\bar{K}N}^2\rangle}$ [fm] &$2.51$ & $2.40$  &$2.27$  \\
  $\sqrt{\langle r_N^{2}\rangle}$ [fm] &$1.81$ & $1.77$  &$1.66$     \\
  $\sqrt{\langle r_{\bar{K}}^{2}\rangle}$ [fm] &$1.61$ & $1.49$  &$1.43$    \\
  $\langle T\rangle^{K^-}_G$ [MeV] &$115+i10.5$ & $125+i31.1$  &$121+i21.1 $    \\
  $\langle V\rangle^{K^-}_G$ [MeV] &$-129-i15.7$ & $-144-i41.5$  &$-137-i37.7$    \\
  $\langle T\rangle_G^{\bar{K}^0}$ [MeV] &$161+i14.2 $ & $177+i43.4 $  &$173+i30.6 $    \\
  $\langle V\rangle_G^{\bar{K}^0}$ [MeV] &$-188  -i21.3$ & $-211 -i60.9$ &$-208 -i55.4$    \\
  $2\langle V\rangle_G^{K^-\bar{K}^0}$ [MeV] &$-30.7+i0.330$ & $-26.2-i9.77$  &$-52.8-i2.72$    \\
  $P_{\bar{K}N}^{I=0}$  &$0.41$ & $0.41$  &$0.40$    \\
  $P_{\bar{K}N}^{I=1}$  &$0.59$ & $0.59$  &$0.60$    \\
 \hline\hline
\end{tabular}  }
\label{tab:Kpppnnn} 
\end{center}
\end{table*}
\begin{table*}[tb]
\begin{center}
\caption{Properties of the calculation for $^6_{\bar{K}}\text{Li}$ system
 with $J^\pi=1^-$.}
{\tabcolsep = 2.7mm
  \begin{tabular}{cccccccccc} \hline\hline
   \multicolumn{4}{c}{$^6_{\bar{K}}\text{Li}~(J^\pi=1^-)$} \\\hline
   Model& \multicolumn{2}{c}{SIDDHARTA}&AY\\
   & Type I  & Type II &  \\\hline
  $B$ [MeV]     &$70.8$ & $77.5$  &$92.9$    \\
  $\Gamma$ [MeV] &$26.4$ & $75.2$ &$88.0$       \\
  $\delta \sqrt{s}$ [MeV]     &$-70.2-i21.5$ & $-13.2-i10.4$   &  \\
  $P_{K^-}$ &$0.07$ & $0.06$  &$0.16$      \\
  $P_{\bar{K}^0}$ &$0.93$ & $0.94$  &$0.84$      \\
  $\sqrt{\langle r_{NN}^2\rangle}$ [fm] &$2.95$ & $2.90$  &$2.83$  \\
  $\sqrt{\langle r_{\bar{K}N}^2\rangle}$ [fm] &$2.54$ & $2.45$  &$2.37$  \\
  $\sqrt{\langle r_N^{2}\rangle}$ [fm] &$1.91$ & $1.88$  &$1.83$     \\
  $\sqrt{\langle r_{\bar{K}}^{2}\rangle}$ [fm] &$1.54$ & $1.45$  &$1.40$    \\
  $\langle T\rangle^{K^-}_G$ [MeV] &$33.3+i5.24$ & $27.3+i16.9$  &$57.1+i8.22$    \\
  $\langle V\rangle^{K^-}_G$ [MeV] &$-26.4-i5.55$ & $-21.1-i16.8$  &$-50.3-i13.7$    \\
  $\langle T\rangle_G^{\bar{K}^0}$ [MeV] &$240+i24.7$ & $268+i61.3$  &$219+i46.1$    \\
  $\langle V\rangle_G^{\bar{K}^0}$ [MeV] &$-294 -i35.2$ & $-332-i89.9$  &$-275-i81.3$    \\
  $2\langle V\rangle_G^{K^-\bar{K}^0}$ [MeV] &$-23.8-i2.40 $ & $-19.2-i9.03$  &$-43.3-i3.31$    \\
  $P_{\bar{K}N}^{I=0}$  &$0.27$ & $0.27$  &$0.29$    \\
  $P_{\bar{K}N}^{I=1}$  &$0.73$ & $0.73$  &$0.71$    \\
 \hline\hline
\end{tabular}  }
\label{tab:Kppppnn1}
\end{center}
\end{table*}
\begin{table*}[tb]
\begin{center}
\caption{Properties of the calculation for $^6_{\bar{K}}\text{He}$ system
 with $J^\pi=1^-$.}
{\tabcolsep = 2.7mm
  \begin{tabular}{cccccccccc} \hline\hline
   \multicolumn{4}{c}{$^6_{\bar{K}}\text{He}~(J^\pi=1^-)$} \\\hline
   Model& \multicolumn{2}{c}{SIDDHARTA}&AY\\
   & Type I  & Type II &  \\\hline
  $B$ [MeV]     &$72.8$ & $80.7$  &$95.6$    \\
  $\Gamma$ [MeV] &$26.0$ & $75.6$ &$88.5$       \\
  $\delta \sqrt{s}$ [MeV]     &$-71.6-i20.8$ & $-13.8-i10.4$   &  \\
  $P_{K^-}$ &$0.93$ & $0.94$  &$0.86$      \\
  $P_{\bar{K}^0}$ &$0.07$ & $0.06$  &$0.14$      \\
  $\sqrt{\langle r_{NN}^2\rangle}$ [fm] &$2.95$ & $2.89$  &$2.81$  \\
  $\sqrt{\langle r_{\bar{K}N}^2\rangle}$ [fm] &$2.53$ & $2.44$  &$2.36$  \\
  $\sqrt{\langle r_N^{2}\rangle}$ [fm] &$1.91$ & $1.87$  &$1.82$     \\
  $\sqrt{\langle r_{\bar{K}}^{2}\rangle}$ [fm] &$1.53$ & $1.44$  &$1.39$    \\
  $\langle T\rangle^{K^-}_G$ [MeV] &$242+i25.0$ & $272+i61.9$  &$223+i46.8$    \\
  $\langle V\rangle^{K^-}_G$ [MeV] &$-298-i35.9$ & $-339-i91.3$  &$-284-i83.0$    \\
  $\langle T\rangle_G^{\bar{K}^0}$ [MeV] &$31.2+i3.96$ & $25.6+i15.4$  &$53.5+i7.40$    \\
  $\langle V\rangle_G^{\bar{K}^0}$ [MeV] &$-24.2 -i4.36$ & $-19.7-i15.2$  &$-46.5-i12.4$    \\
  $2\langle V\rangle_G^{K^-\bar{K}^0}$ [MeV] &$-23.3-i1.69$ & $-18.8-i8.64$ &$-42.3-i3.10$    \\
  $P_{\bar{K}N}^{I=0}$  &$0.27$ & $0.26$  &$0.29$    \\
  $P_{\bar{K}N}^{I=1}$  &$0.73$ & $0.74$  &$0.71$    \\
 \hline\hline
\end{tabular}  }
\label{tab:Kpppnnn1} 
\end{center}
\end{table*}
\begin{figure*}[tbh]
 \includegraphics[width=0.48\textwidth,clip]{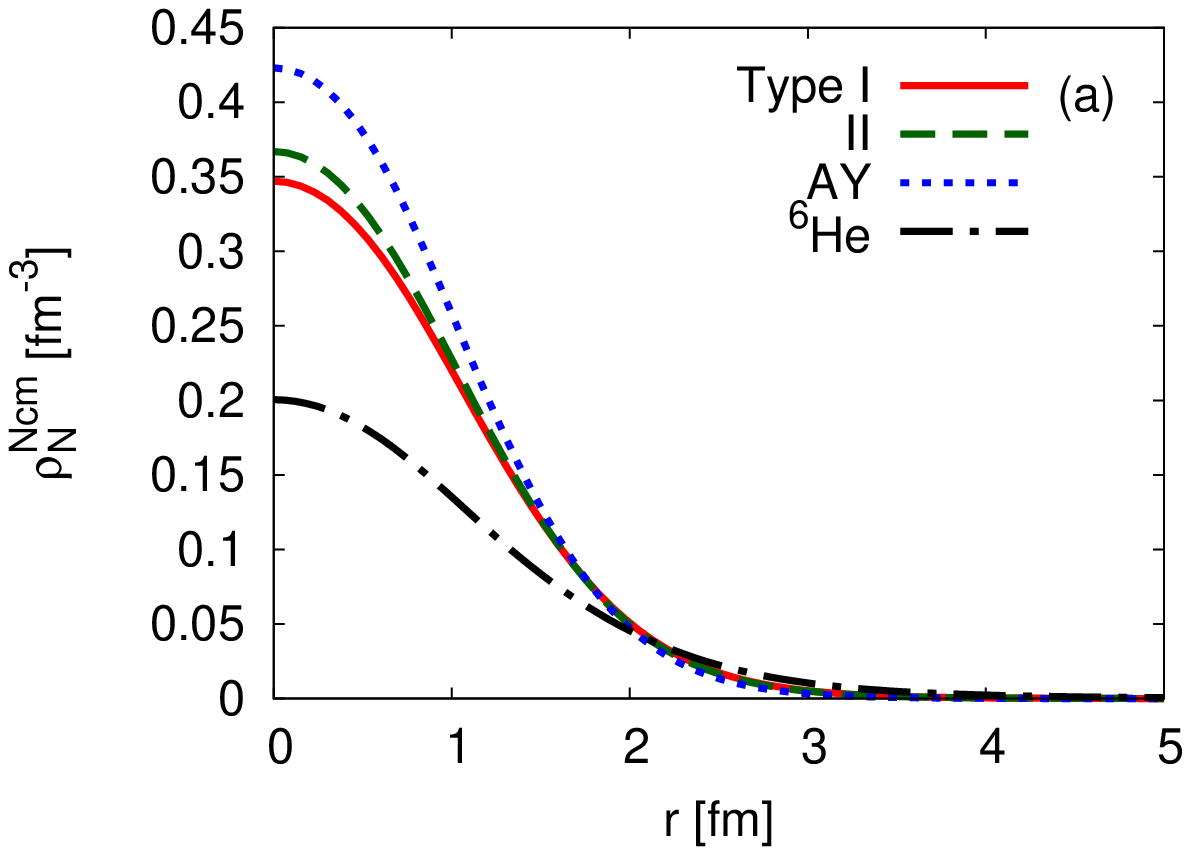}
 \includegraphics[width=0.48\textwidth,clip]{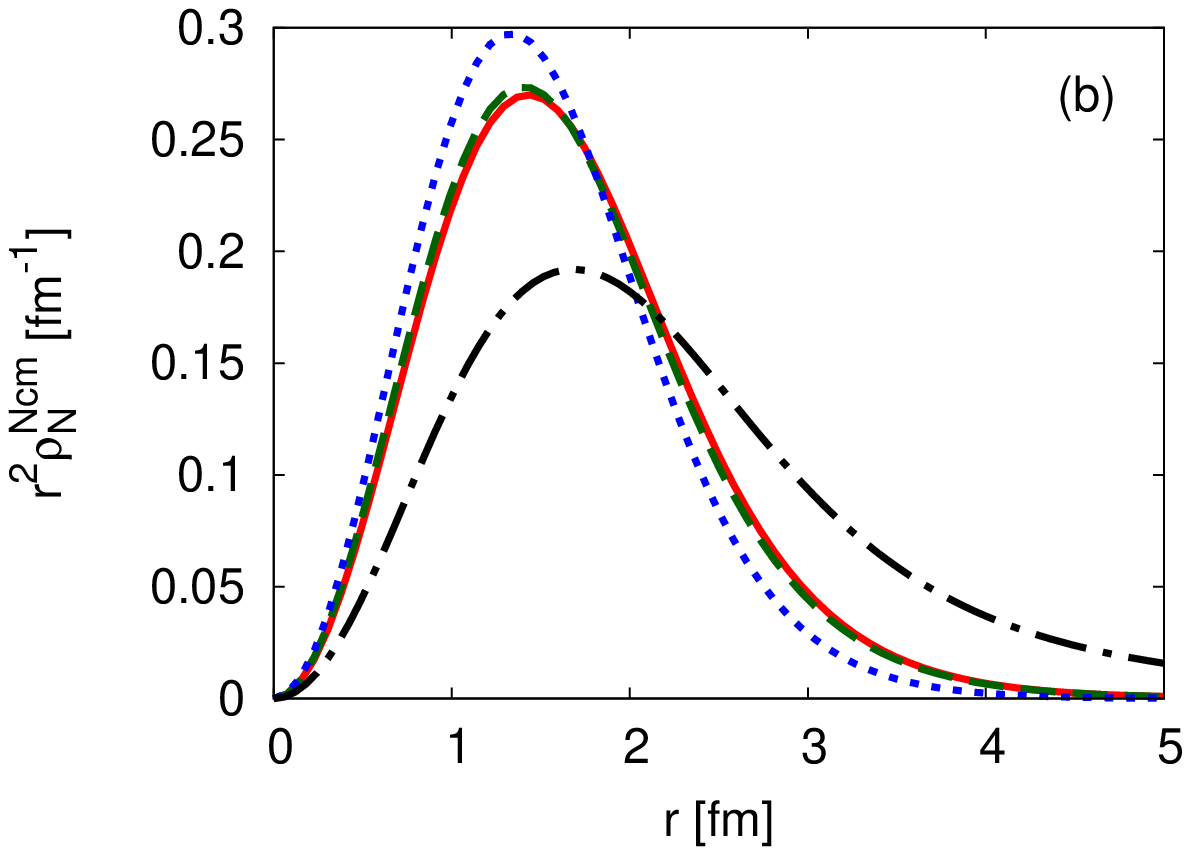}
 \caption{(Color online) Same as Fig.~\ref{fig:K-pp_N_dist}
 but for  $^6_{\bar{K}}\text{He}$ system with
 $J^\pi=0^-$.
 The nucleon density
 distributions for $^6$He with $J^\pi=0^+$ is plotted for comparison.}
 \label{fig:K-pppnnn_N_dist}
\end{figure*}
\begin{figure*}[tbh]
 \includegraphics[width=0.48\textwidth,clip]{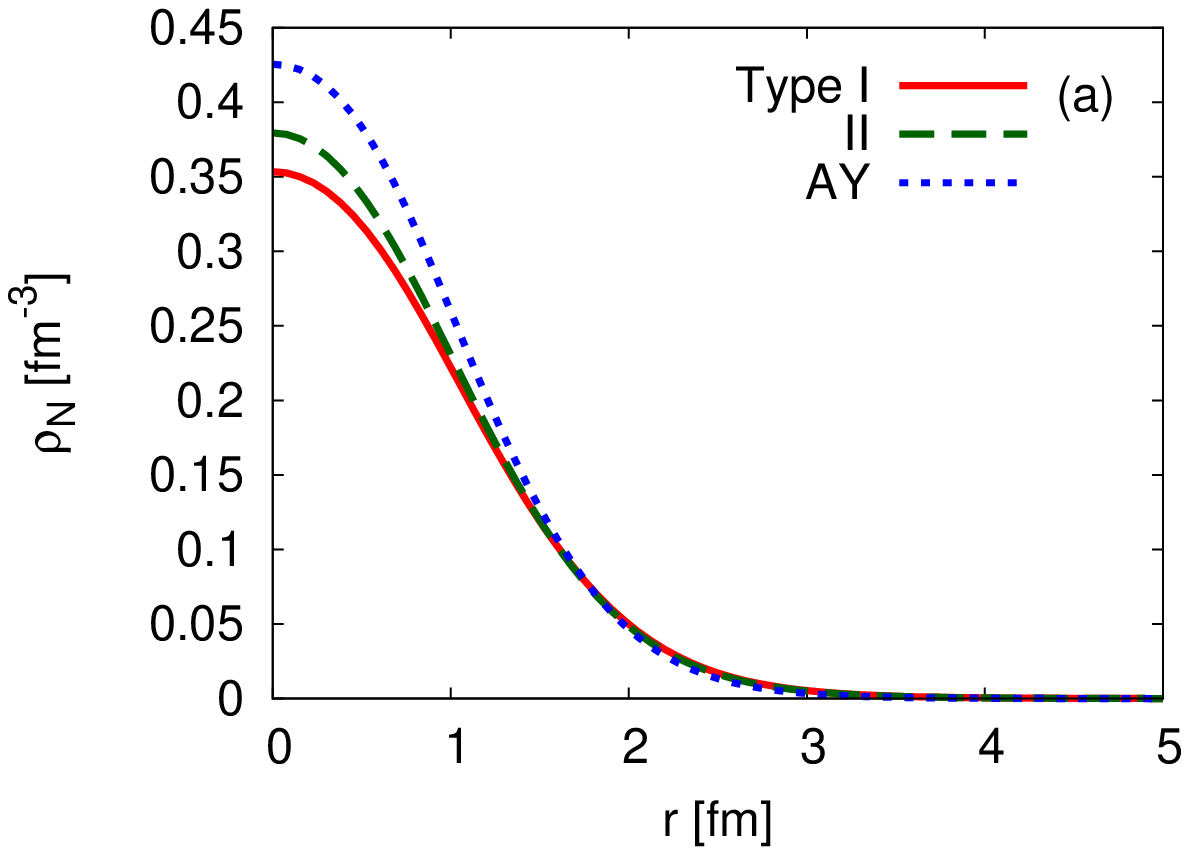}
 \includegraphics[width=0.48\textwidth,clip]{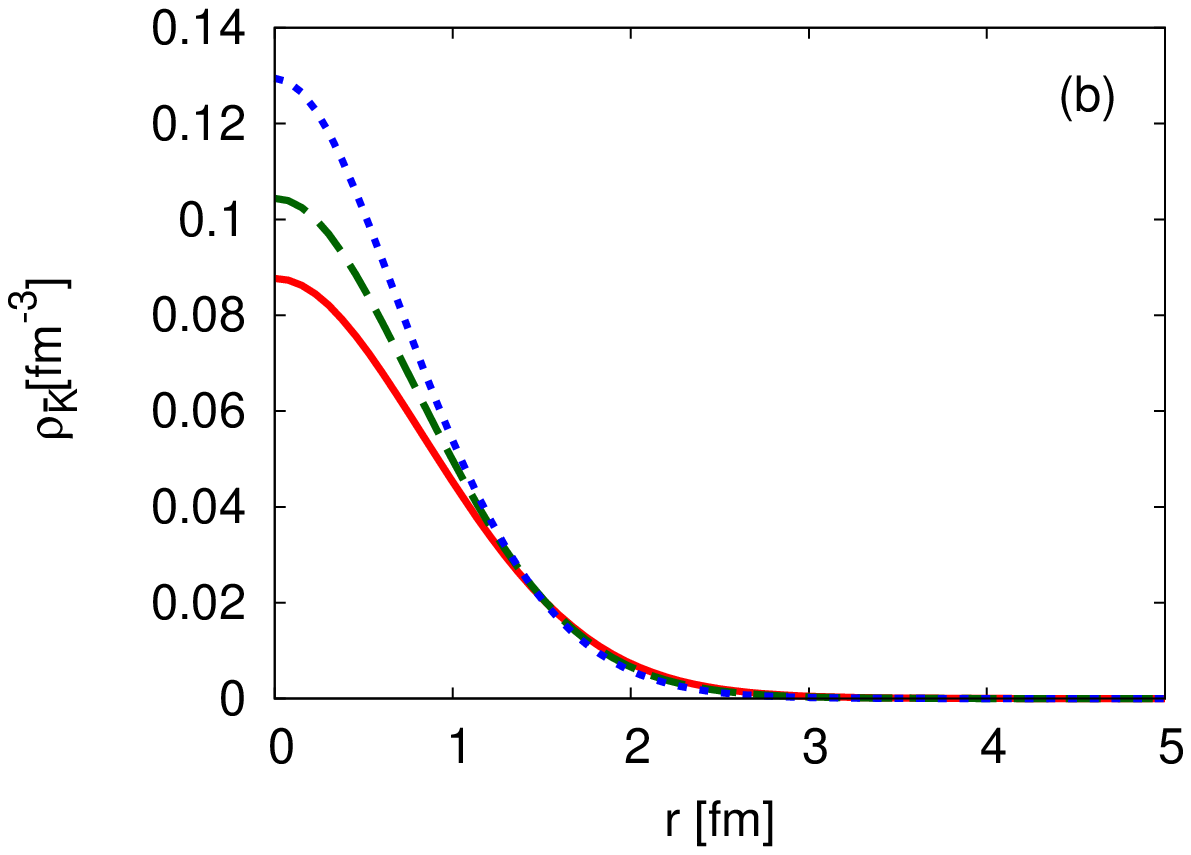}
 \caption{(Color online) Same as Fig.~\ref{fig:K-pp_dist}
 but for  $^6_{\bar{K}}\text{He}$ system with
 $J^\pi=0^-$.
}
 \label{fig:K-pppnnn_dist}
\end{figure*}
\begin{figure*}[tbh]
 \includegraphics[width=0.48\textwidth,clip]{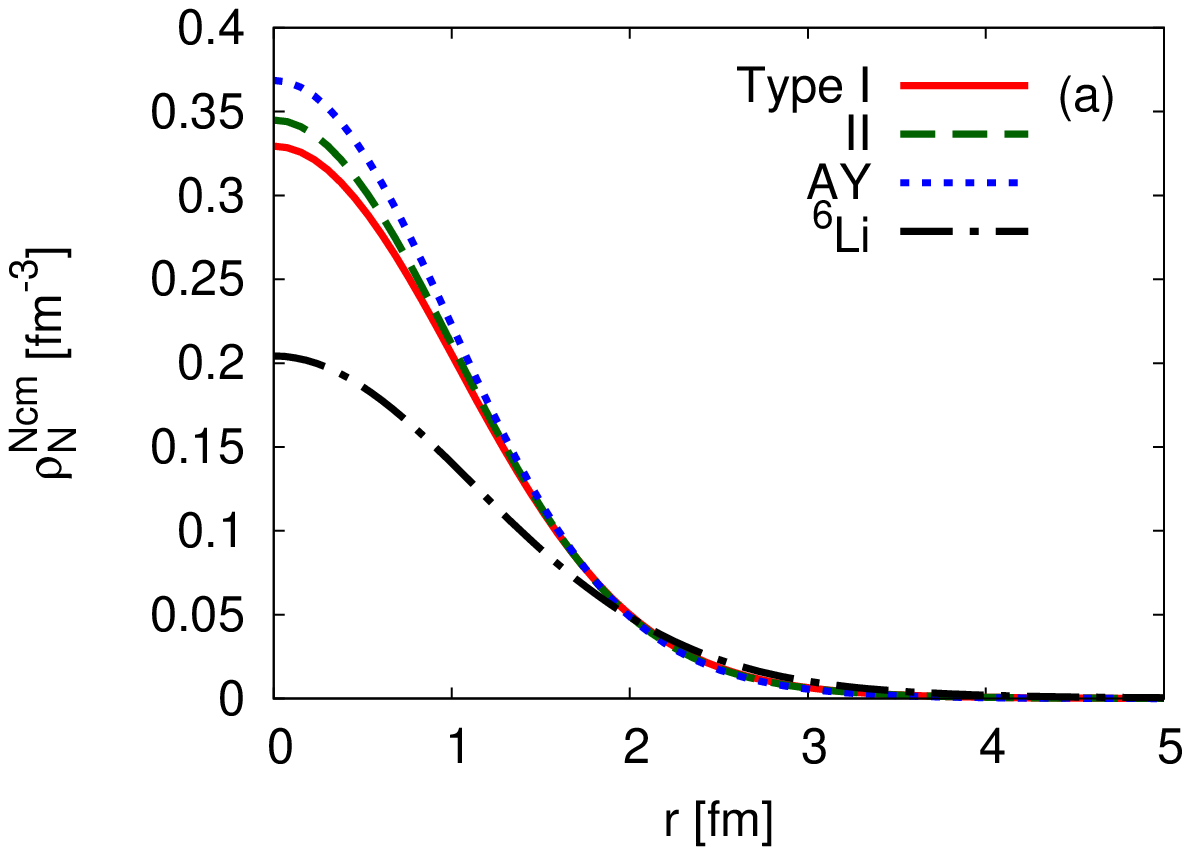}
 \includegraphics[width=0.48\textwidth,clip]{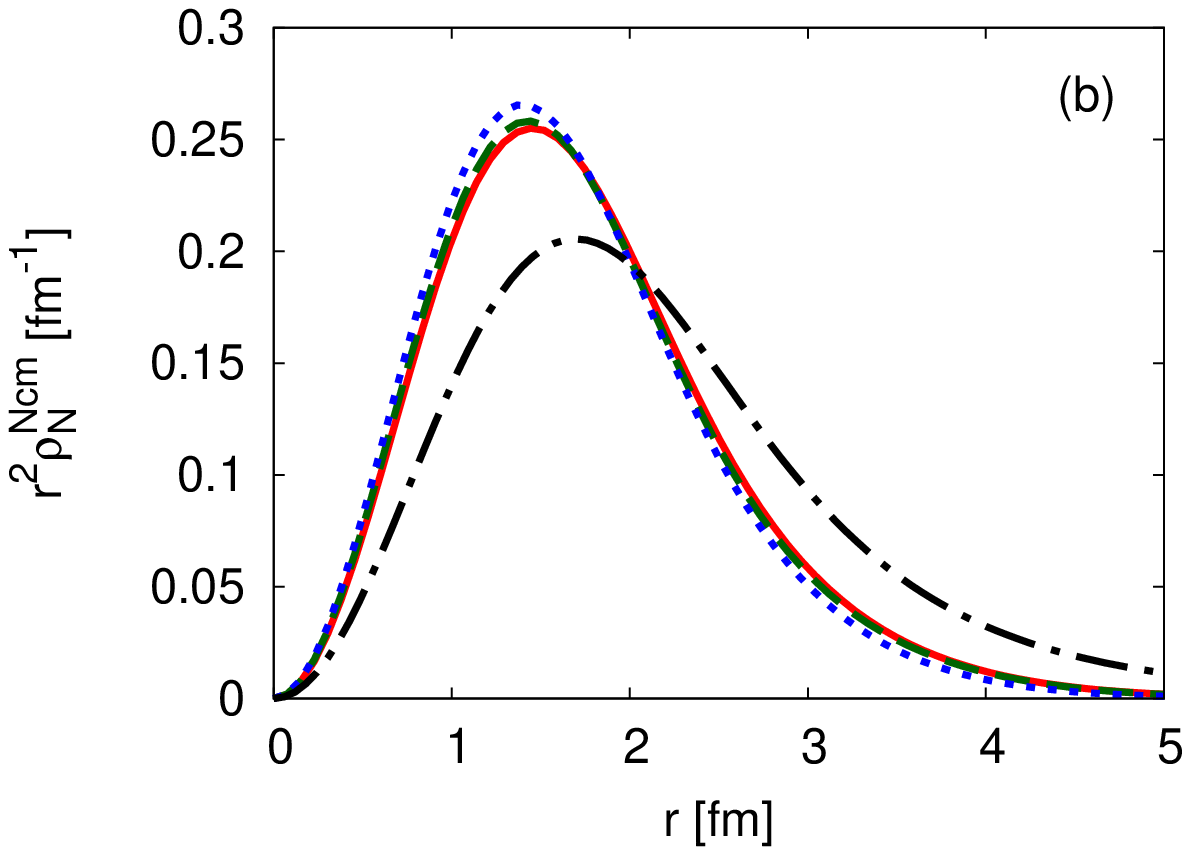}
 \caption{(Color online) Same as Fig.~\ref{fig:K-pp_N_dist}
 but for  $^6_{\bar{K}}\text{He}$ system with
 $J^\pi=1^-$.
 The nucleon density
 distributions for $^6$Li with $J^\pi=1^+$ is plotted for comparison.}
 \label{fig:K-pppnnn1_N_dist}
\end{figure*}
\begin{figure*}[tbh]
 \includegraphics[width=0.48\textwidth,clip]{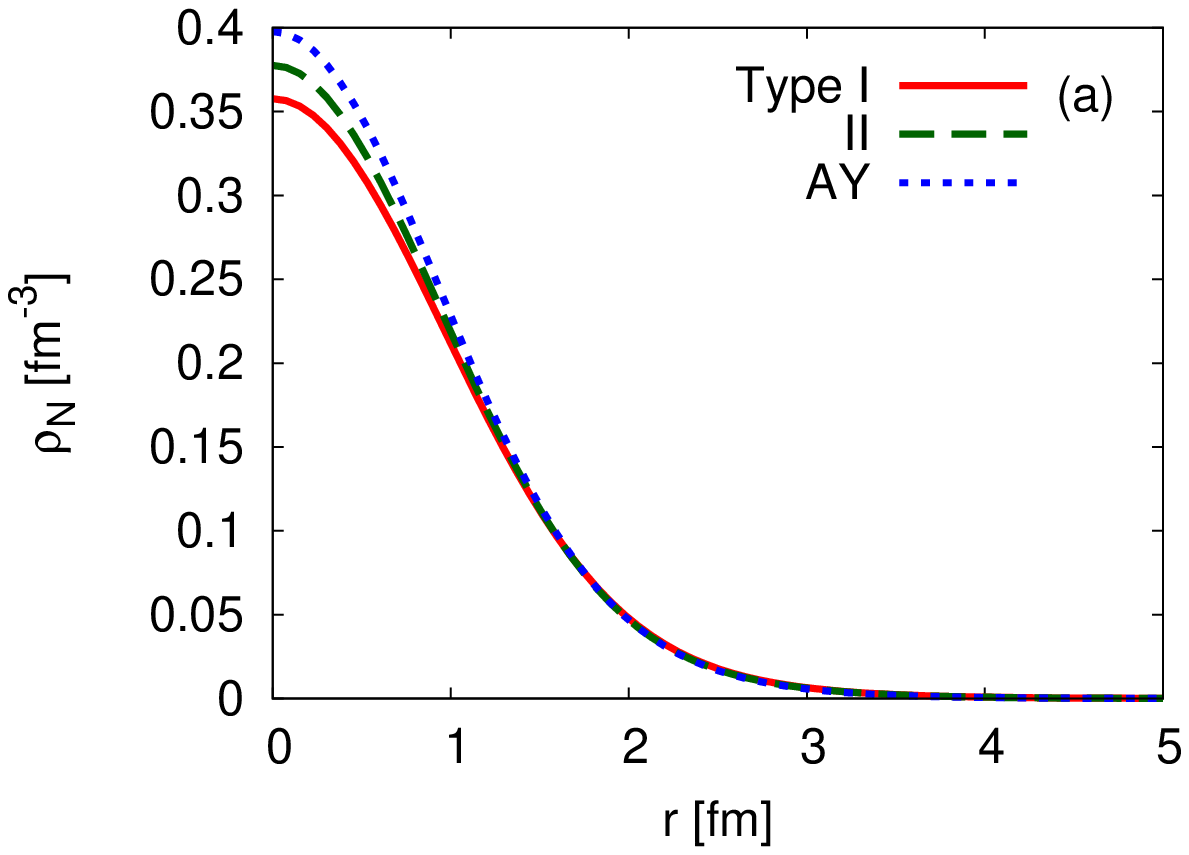}
 \includegraphics[width=0.48\textwidth,clip]{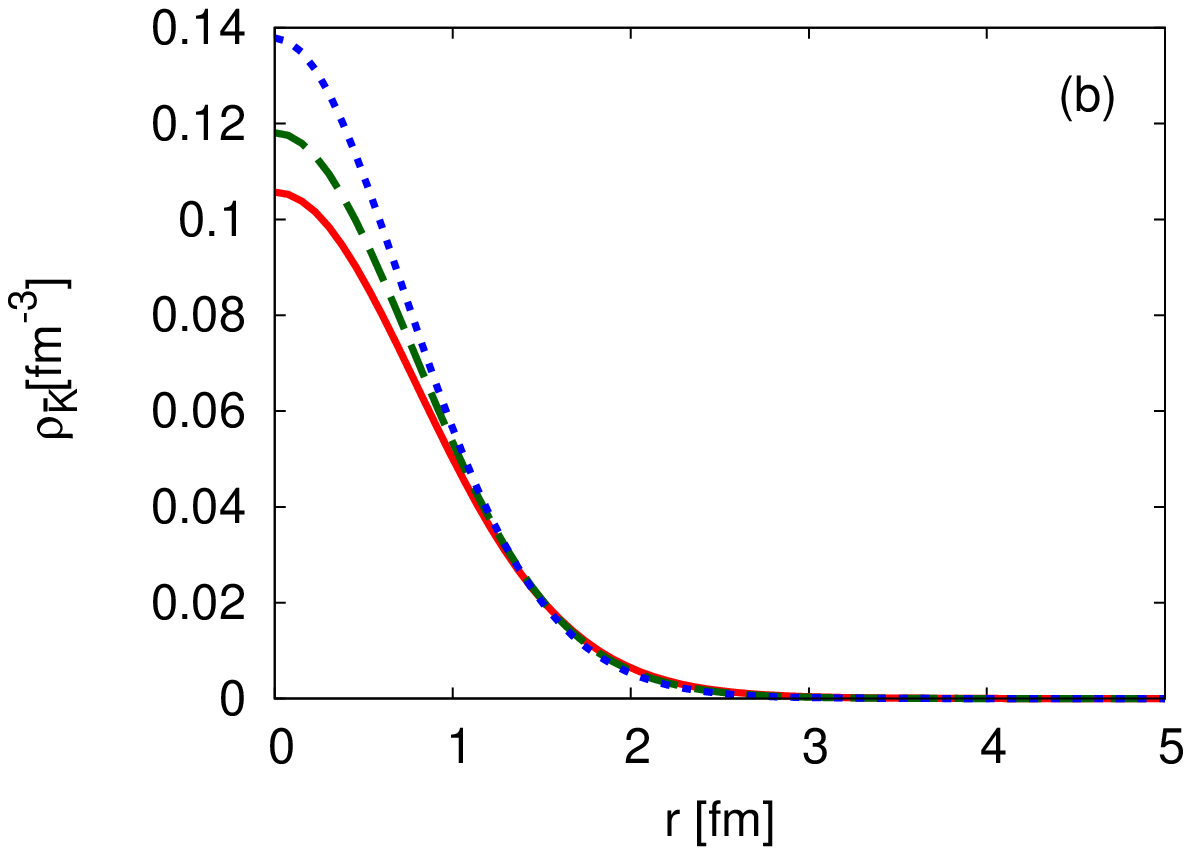}
 \caption{(Color online) Same as Fig.~\ref{fig:K-pp_N_dist}
 but for  $^6_{\bar{K}}\text{He}$ system with
 $J^\pi=1^-$.
}
 \label{fig:K-pppnnn1_dist}
\end{figure*}

Finally, we investigate the structure of the seven-body systems,
strange hexabaryon $\bar{K}NNNNNN$.
The ground state of the six-nucleon systems without an antikaon
is $^6$Li with $J^\pi=1^+$,
and the $J^\pi=0^+$ isospin-triplet states,
$^6$He, $^6$Li and $^6$Be, are the excited states.
By adding an antikaon, we can construct two isospin doublets with
$J^{\pi}=1^{-}$ and $0^{-}$. We find quasi-bound states in these quantum
numbers, while we do not find any states below the $(\bar{K}NNNN)+2N$
threshold for $I=3/2$ with $J^\pi=0^-$, such as
$K^-ppnnnn$-$\bar{K}^0pnnnnn$ ($\equiv~^6_{\bar{K}}\text{H}$) system.

Tables~\ref{tab:Kppppnn},
\ref{tab:Kpppnnn},
\ref{tab:Kppppnn1},
and \ref{tab:Kpppnnn1}
list our results of 
$K^-ppppnn$-$\bar{K}^0pppnnn$ ($\equiv~^6_{\bar{K}}\text{Li}$) system with $J^\pi=0^-$,
$K^-pppnnn$-$\bar{K}^0ppnnnn$ ($\equiv~^6_{\bar{K}}\text{He}$) system with $J^\pi=0^-$,
$^6_{\bar{K}}\text{Li}$ system with $J^\pi=1^-$,
and $^6_{\bar{K}}\text{He}$ system with $J^\pi=1^-$, respectively.
The binding energies for the $J^\pi=0^-$ ($1^-$) states with 
Type II is 9-10 MeV (7-8 MeV) larger
than those with Type I.
The decay widths with Type II are three times larger than
those with Type I. The rms distances
$\sqrt{\langle r_{NN}^2 \rangle}$, $\sqrt{\langle r_{\bar{K}N}^2 \rangle}$,
$\sqrt{\langle r_{N}^2 \rangle}$, and
$\sqrt{\langle r_{K}^2 \rangle}$ with 
Type II are slightly smaller than
those with Type I. 
The probabilities $P_{K^-}$ and
$P_{\bar{K}^0}$ are not sensitive to the choice of Types I and II. 
As in the case of
four- and five-body systems, the diagonal channels give important contributions to 
the decay width.
In contrast, 
both of the diagonal and the off-diagonal components
produce about a half of the binding energy. 
The diagonal components of the channel with $^6\text{Li}$,
that is, the $K^{-}$ channel in $^6_{\bar{K}}\text{He}$ and the $\bar{K}^{0}$ channel in $^6_{\bar{K}}\text{Li}$, are
important especially for the $J^\pi=1^-$ states.

The dominant component of the $^6_{\bar{K}}\text{Li}$
($^6_{\bar{K}}\text{He}$) system with $J^\pi=0^-$ is the $K^-ppppnn$ ($\bar{K}^0ppnnnn$) channel,
while that of the $^6_{\bar{K}}\text{Li}$
($^6_{\bar{K}}\text{He}$) system with $J^\pi=1^-$
is the $\bar{K}^0pppnnn$ ($K^-pppnnn$) channel.
For the spin-singlet states ($J^\pi=0^-$), the core nuclei in 
$^6_{\bar{K}}\text{Li}$ and 
$^6_{\bar{K}}\text{He}$
are the isospin-triplet states of $^6$Be, $^6$Li and $^6$He,
and the channels with larger fraction of the
$\bar{K}N$ $I=0$ components are favored.
Meanwhile, for the spin-triplet states ($J^\pi=1^-$),
the core nucleus with $J^\pi=1^+$ ($\sim ^6$Li) is the
isospin-singlet state. 
The spin-triplet $^6$Li is the ground
state of the six-nucleon systems,
while $^6$Be and $^6$He with
$J^\pi=1^+$ are not bound.
Therefore, the nucleons in the $\bar{K}^0pppnnn$ ($K^-pppnnn$)
channel feel larger attraction 
than that in the other channel. This determines the dominant component
in the $J^\pi=1^-$ state.
This is also the reason why the $\bar{K}^0$ ($K^-$) diagonal component
gains the large binding energy in the spin-singlet
$^6_{\bar{K}}\text{Li}$ ($^6_{\bar{K}}\text{He}$) system as
similar to the $\bar{K}NNNN$ system.

The Coulomb splitting between $^6_{\bar{K}}\text{Li}$ and
$^6_{\bar{K}}\text{He}$ in the $J^\pi=0^-$ channel ($0.3$-$0.8$ MeV) is smaller than
the splitting in $J^\pi=1^-$ ($2.0$-$3.2$ MeV).
In the dominant $K^-ppppnn$ component of the $^6_{\bar{K}}\text{Li}$ with $J^\pi=0^-$, 
there are four attractive and six repulsive
Coulombic pairs,
while the dominant $\bar{K}^0ppnnnn$ channel in
$^6_{\bar{K}}\text{He}$ contains one repulsive
pair.
Therefore, the Coulomb interaction in $^6_{\bar{K}}\text{Li}$
system is expected to be slightly repulsive than $^6_{\bar{K}}\text{He}$ system.
On the other hand, in 
the dominant $\bar{K}^0pppnnn$ component of the $^6_{\bar{K}}\text{Li}$ system with $J^\pi=1^-$, 
there are three repulsive
Coulombic pairs, and the dominant $K^-pppnnn$ channel in
$^6_{\bar{K}}\text{He}$ has 
three attractive and
three repulsive pairs. 
In the $^6_{\bar{K}}\text{He}$ system, the rms distance
$\sqrt{\langle r^2_{\bar{K}N}\rangle}$ is smaller than
$\sqrt{\langle r^2_{NN}\rangle}$, and therefore, the $\bar{K}N$ Coulomb attraction
is stronger than the $NN$ 
repulsion, and the Coulomb interaction %
works in total attractively
in the $K^-pppnnn$ channel.
Therefore, the Coulomb splitting between $^6_{\bar{K}}\text{Li}$ and
$^6_{\bar{K}}\text{He}$ in $J^\pi=1^-$ becomes larger than
the splitting in $J^\pi=0^-$.

Next, we compare the spin-singlet and triplet states.
With Type I, the binding energy of
$^6_{\bar{K}}\text{Li}$ with 
$J^\pi=1^-$ is 1 MeV larger than the 
$J^\pi=0^-$ state, and $^6_{\bar{K}}\text{He}$ with 
$J^\pi=1^-$ is 2.2 MeV larger than in
$J^\pi=0^-$.
With Type II, 
the binding energy of
$^6_{\bar{K}}\text{Li}$ with  
$J^\pi=1^-$ is 2.2 MeV smaller than the
$J^\pi=0^-$ state, and $^6_{\bar{K}}\text{He}$ with
$J^\pi=1^-$ is 0.8 MeV larger than the $J^\pi=0^-$ state.
Except for $^6_{\bar{K}}\text{Li}$ with Type II,
the binding energies of the spin-triplet states are larger than the spin-singlet states. This is in accordance with the level structure of
the six-nucleon systems.
With Type II, the magnitude of the real part of 
$\delta\sqrt{s}$ used
in the two-body $\bar{K}N$
interaction is smaller, and the $\bar{K}N$
interaction becomes more attractive than that with Type I. 
Because the $J^\pi=0^-$ state contains larger fraction of the
$I=0$ $\bar{K}N$ components than the $J^\pi=1^-$ state, as in the case of the strange dibaryon
$\bar{K}NN$ systems, 
the $\bar{K}N$ interaction with 
Type II is so
strong that the binding energy of 
$^6_{\bar{K}}\text{Li}$ with 
$J^\pi=0^-$ becomes larger than the $J^\pi=1^-$ state.
In other words, by adding an antikaon, the spin of the ground state of the six-nucleon system may change depending on the strength of the $\bar{K}N$ interaction. The inversion of the level structure of the ground state and the first excited state by the antikaon
is also seen in the two-nucleon systems (strange dibaryon $\bar{K}NN$). The ground state of the two-nucleon sector without the antikaon is the spin-triplet deuteron, while the spin-singlet channel is unbound.
As discussed in Sec.~\ref{subsec:KNN}, by injecting an antikaon, the ground state is spin singlet which maximizes the fraction of the $\bar{K}N(I=0)$ component. Recalling that the $^6$He and $^6$Li are well approximated
by an $\alpha+N+N$ three-body model (See, for example, Ref.~\cite{horiuchi07}
and references therein), 
the difference of $J^\pi=1^-$ 
and $J^\pi=0^-$ states of seven-body systems
can be essentially caused by the 
difference of the $\bar{K}NN$ subsystems. Similar to the three-body systems,
the level inversion can take place also in the seven-body systems,
which is driven by the balance between the nuclear structure and the attraction in the $\bar{K}N$ system.

This property is more pronounced if the strength of the $\bar{K}N$ attraction is further increased. For instance, when we use the AY potential which is
more attractive than the SIDDHARTA potential,
the binding energies of $^6_{\bar{K}}\text{Li}$ and
$^6_{\bar{K}}\text{He}$ with 
$J^\pi=0^-$ become $10$ and 7 MeV larger than
those with $J^\pi=1^-$.
Namely, the stronger $\bar{K}N$ attraction leads to more drastic level inversion in the seven-body systems. In this way,
there is a possibility to extract the information on the $\bar{K}N$
interaction not only from the binding energies but also from the ground state quantum
number $J^\pi$ and from the splitting between $J^\pi=0^-$ and $1^-$ states.

\begin{figure*}[tbh]
 \includegraphics[width=0.48\textwidth,clip]{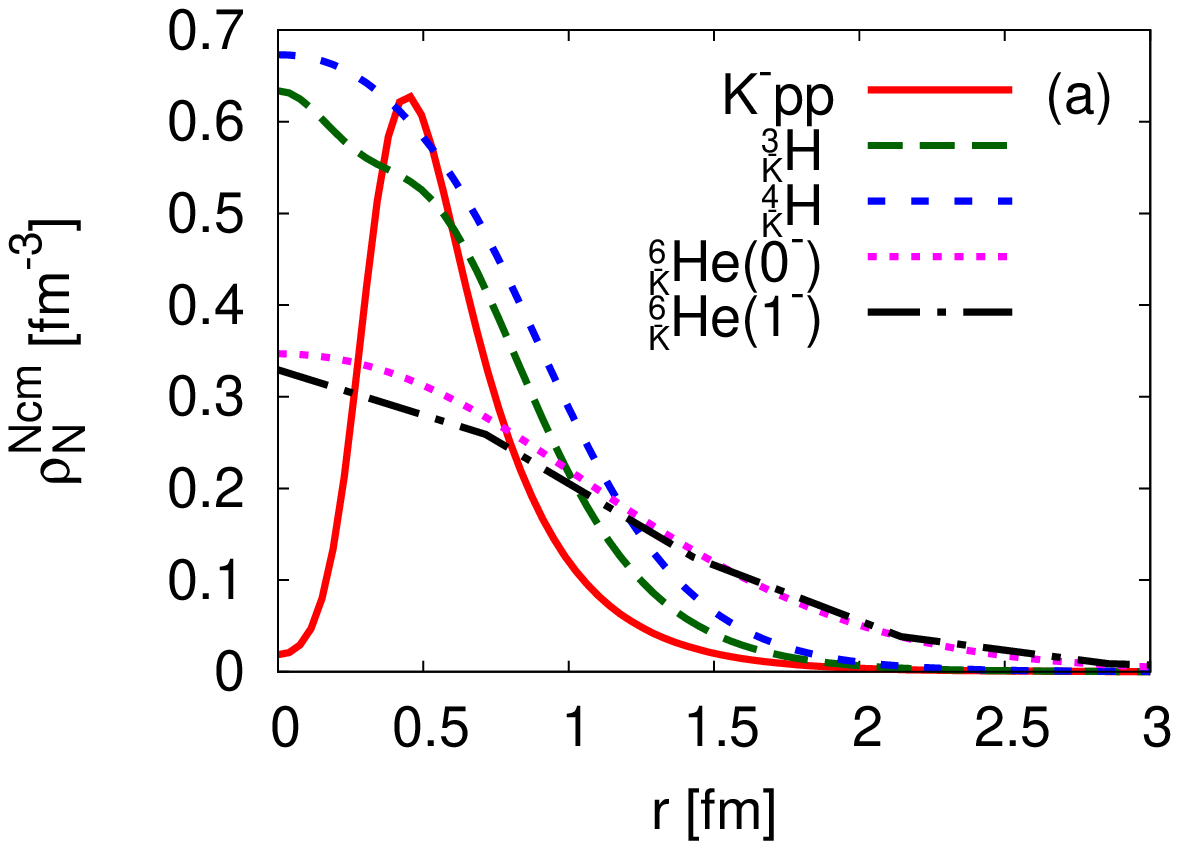}
 \includegraphics[width=0.48\textwidth,clip]{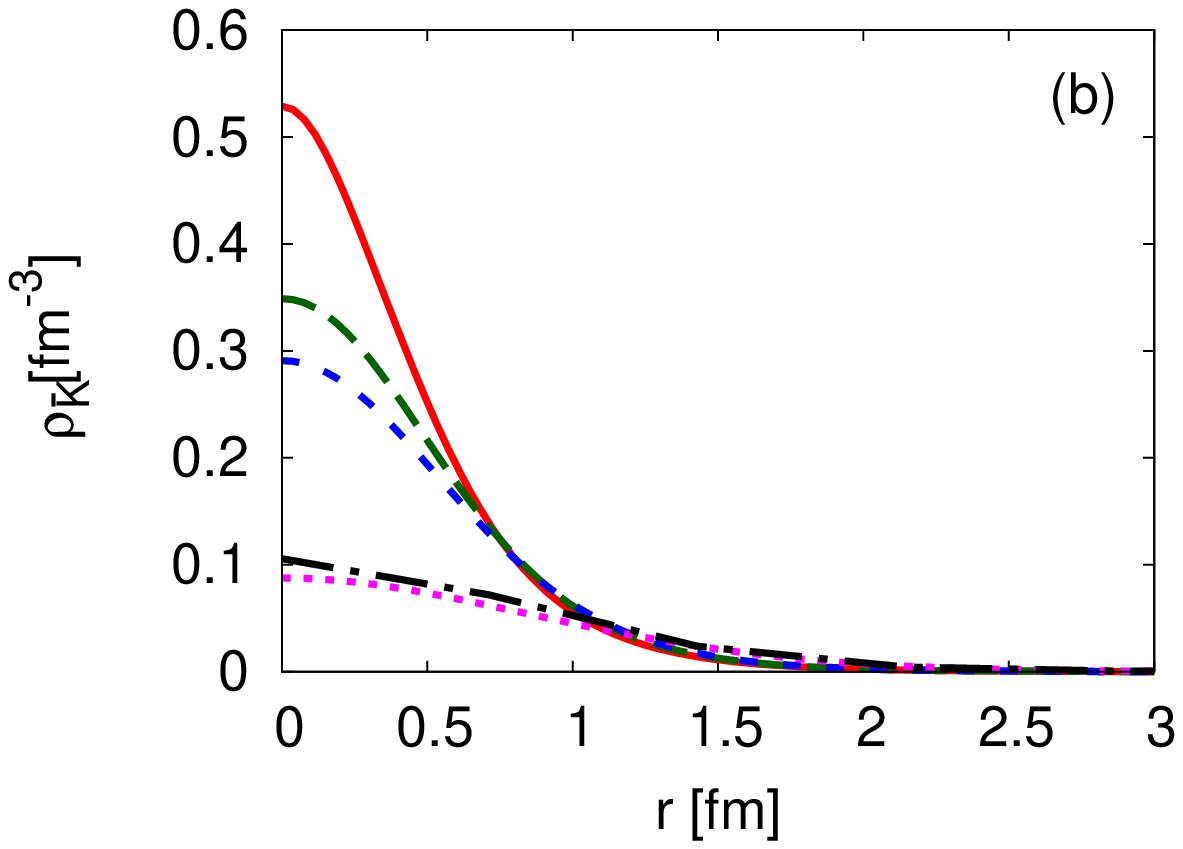}
 \caption{(Color online) (a) Nucleon density distributions 
   $\rho_N^{N\text{cm}}(r)$ and (b) antikaon density distributions 
   $\rho_{\bar{K}}(r)$
     for the three-, four-, five-, and seven-body kaonic nuclear systems.
 SIDDHARTA potential Type I is employed for $\bar{K}N$ interaction.
}
 \label{fig:N_dist}
\end{figure*}

Finally, we show, in Figs.~\ref{fig:K-pppnnn_N_dist},
  ~\ref{fig:K-pppnnn_dist},~\ref{fig:K-pppnnn1_N_dist}, 
  and \ref{fig:K-pppnnn1_dist},
  the particle density distributions of 
  the
  $^6_{\bar{K}}\text{He}$ 
  system with 
  $J^\pi=0^-$ and $1^-$.
  The nucleon density distributions of 
  the $^6$He ($J^\pi=0^+$) and $^6$Li ($J^\pi=1^+$) systems are also plotted for comparison.
In the $\bar{K}NNNNNN$ system with 
$J^\pi=0^-$, the central nucleon
density becomes slightly larger than that with 
$J^\pi=1^-$
and about two times larger than that in the $^6$Li.
Meanwhile, the central antikaon density with 
$J^\pi=0^-$ is slightly smaller than that
with 
$J^\pi=1^-$.

In Fig.~\ref{fig:N_dist}, we summarize the nucleon and antikaon density
distributions of various kaonic nuclei from three- to seven-body systems.
The central nucleon densities of the seven-body systems become
about a half,
and antikaon densities become one third of the densities of the
four- and five-body systems.
The central nucleon
  densities of the five-body systems are highest in the light kaonic
  nuclei up to seven-body systems,
  while the densities are not as high as those suggested by using the
  effective interaction based on  $g$-matrix approach in
  Refs.~\cite{Dote:2002db,Dote:2003ac}.
The large nucleon density in the five-body system is mainly caused by the formation of an $\alpha$-particle configuration, rather than maximizing the $\bar{K}N(I=0)$ pairs.
  For the antikaon distribution, the central densities become smaller as the number of nucleons increases.
Because the antikaon feels attraction from all the nucleons, its spatial extent increases in a large nucleus.

\section{Summary}
\label{sec:summary}
We have studied structure of the light kaonic nuclei,
$\bar{K}NN$, $\bar{K}NNN$, $\bar{K}NNNN$, and $\bar{K}NNNNNN$
with a powerful few-body approach, the correlated Gaussian (CG) method.
Fully converged three- to seven-body solutions are obtained by
the stochastic variational method (SVM). As a realistic $\bar{K}N$ interaction, we employ the SIDDHARTA potential constructed based on the NLO chiral SU(3) dynamics with the SIDDHARTA constraint obtained from
Refs.~\cite{Bazzi:2011zj,Bazzi:2012eq}.

We find one quasi-bound state in the $\bar{K}NN$, $\bar{K}NNN$, and $\bar{K}NNNN$ systems, and two quasi-bound states with $J^{\pi}=0^{-}$ and $1^{-}$ in the $\bar{K}NNNNNN$ system. All the states are found above the $\pi\Sigma$ emission threshold.
The central densities of nucleons are enhanced by an injected antikaon,
and become about two times larger than those without an antikaon. The
central nucleon density reaches its maximum in the $\bar{K}NNNN$ system
with $J^\pi=0^-$,
where the nucleons can form an $\alpha$-particle configuration. The rms radius of the antikaon increases along with the nucleon rms radius when the mass number is increased.

By decomposing the eigenenergy into different contributions, we find that the $K^-$-$\bar{K}^0$ channel coupling is important for the binding
of the light kaonic nuclei.
For the $\bar{K}NN$, $\bar{K}NNN$ and $\bar{K}NNNNNN$ with $J^\pi=0^-$,
the core nuclei belong to same isospin multiplet,
and the mixing between $K^-$ and $\bar{K}^0$ channels and
the energy gains from the off-diagonal components are large.
Meanwhile, for the $\bar{K}NNNN$ and $\bar{K}NNNNNN$ with $J^\pi=1^-$,
the channel with core nucleus $^4$He or $^6$Li is dominant,
and the energy gains from both of the off-diagonal and diagonal
components with $^4$He or $^6$Li become large.

To take into account the energy dependence of the SIDDHARTA potential,
we examine two methods to determine the $\bar{K}N$ two-body energy in
$\mathscr{N}$-body systems (Types I and II). Quantitatively, 
the binding energies with Type II become
gradually larger than those with Type I
as the number of particles increase, and 
the decay
widths with Type II become 2-3 times larger than those with Type I.
The qualitative features of the kaonic nuclei are not sensitive to the
choice of the method.

In order to examine the predictions of deeply-bound and high-density kaonic nuclei~\cite{Akaishi:2002bg,Dote:2002db,Dote:2003ac},
we also use the AY potential model.
When we employ the AY potential, the binding energies are about $20$-$30$
MeV larger than those with the SIDDHARTA potential for each system, and decay
widths become around $60$-$80$ MeV.
Even in this case, the obtained binding energies up to
seven-body systems except that $\bar{K}NNNNNN$
with $J^\pi=0^-$ are smaller than $100$ MeV predicted by using
the optical
potential approach~\cite{Akaishi:2002bg}.
The central density of the $\bar{K}NNNN$ system is not as high as those suggested by using effective
$\bar{K}N$ and $NN$ interactions based on the $g$-matrix approach in
Ref.~\cite{Dote:2002db,Dote:2003ac}.

The comparison of the two $\bar{K}N$ potential models (SIDDHARTA and AY) leads to interesting
results in the seven-body systems.
If the $\bar{K}N$ attraction is not so strong,
we see the spin-triplet ground ($J^\pi=1^-$)
and the spin-singlet ($J^\pi=0^-$) excited states
reflecting the lightest core nucleus, $^6$Li with $J^\pi=1^+$.
If the $\bar{K}N$ interaction is strong enough
as the AY potential,
the level ordering of the $J^\pi=0^-$ and $J^\pi=1^-$ states is
inverted.
Therefore, it is possible to extract the information on
the $\bar{K}N$ interaction from the ground state quantum
number $J^\pi$ as well as the energy splitting between $J^\pi=0^-$ and $1^-$
of the seven-body kaonic states.

In this work, we employ the single channel $\bar{K}N$ interaction where
the $\pi\Sigma$ channel coupling effect are renormalized into
  its imaginary part.
This potential model reproduces the two-pole structure of
$\Lambda(1405)$,
while we could not find two-pole structure in the kaonic nuclei.
One of these poles with the large binding energy and width is 
originated from the $\pi\Sigma$ resonance pole.
In order to study the effect of 
the other pole in the kaonic nuclei, it may be
necessary to take into account the channel
coupling effect of $\bar{K}N$-$\pi\Sigma$ explicitly.
The work in this direction is underway and will be
  reported elsewhere.

\begin{acknowledgments}

 The authors thank A. Gal, A. Ohnishi, A. Dot\'e and Y. Ikeda for helpful comments and discussions.
 The numerical calculation has been performed on supercomputers (NEC
 SX-ACE) at the Research Center for Nuclear Physics, Osaka University and
 (CRAY XC40) at the Yukawa Institute for Theoretical Physics, Kyoto
 University.
 This work was partly supported by the Grants-in-Aid for Scientific
 Research on Innovative Areas from MEXT (Grant No. 2404:24105008), by JSPS
KAKENHI Grant No. 24740152, and by the Yukawa
International Program for Quark-Hadron Sciences (YIPQS).
\end{acknowledgments}

\bibliography{biblio}

\end{document}